\documentclass[2p,review,10pt]{elsarticle} 

\usepackage[T1]{fontenc}
\usepackage[utf8]{inputenc}
\usepackage[english]{babel}
\usepackage{graphicx}
\usepackage{wrapfig}
\usepackage[usenames,dvipsnames]{color}
\usepackage{amssymb}
\usepackage{multirow}
\usepackage{amsfonts,amsmath,amssymb}
\usepackage{epsfig}
\usepackage{subfig}
\usepackage{caption}
\usepackage{textcomp}
\usepackage[hang,flushmargin]{footmisc}
\usepackage{mathrsfs}
\usepackage{amsthm}
\usepackage{amsmath}
\usepackage{bbm}
\usepackage{yfonts}
\usepackage[colorlinks=true] {hyperref}
\usepackage{algorithmic}
\usepackage[boxed]{algorithm2e}
\usepackage{mathrsfs}
\usepackage{xcolor}
\usepackage{comment}
\usepackage{csquotes}
\usepackage{soul}
\usepackage{mathtools}
\usepackage[export]{adjustbox} 
\usepackage{booktabs, tabularx}
\usepackage{siunitx}
\usepackage{multirow}
\DeclarePairedDelimiter{\ceil}{\lceil}{\rceil}

\usepackage[a4paper,
            left=0.7in,right=0.7in,top=1in,bottom=1in,%
            footskip=.25in]{geometry}
            
\newcommand{\bepsilon}{\boldsymbol{\epsilon}}
\newcommand{\bsigma}{\boldsymbol{\sigma}}

\newcommand{\bu}{\mathbf{u}}
\newcommand{\bv}{\mathbf{v}}

\newcommand{\gnn}{g_\mathrm{n}}

\newcommand{\gtt}{g_\mathrm{t}}
\newcommand{\utt}{u_\mathrm{t}}
\newcommand{\unn}{u_\mathrm{n}}
\newcommand{\pnn}{p_\mathrm{n}}
\newcommand{\ptt}{p_\mathrm{t}}

\newcommand{\gttd}{\dot{g}_\mathrm{t}}

\begin{document}
\begin{frontmatter}
\title{A framework for the analysis of fully coupled normal and tangential contact problems with complex interfaces}
\date{}

\author[1]{Jacopo Bonari\corref{cor1}}
\ead{jacopo.bonari@imtlucca.it}

\author[1]{Marco Paggi}
\ead{marco.paggi@imtlucca.it}

\author[2]{José Reinoso}
\ead{jreinoso@us.es}

\address[1]{IMT School for Advanced Studies Lucca, Piazza San Francesco 19, 55100 Lucca, Italy}
\address[2]{Elasticity and Strength of Materials Group, School of Engineering, Universidad de Sevilla, Camino de los Descubrimientos s/n, 41092, Seville, Spain}

\cortext[cor1]{Corresponding author}
\journal{Finite Elements in Analysis and Design}

\begin{abstract}
An extension to the interface finite element with eMbedded Profile for Joint Roughness (MPJR interface finite element) is herein proposed for solving the frictional contact problem between a rigid indenter of any complex shape and an elastic body under generic oblique load histories. The actual shape of the indenter is accounted for as correction of the gap function. A regularised version of the Coulomb friction law is employed for modeling the tangential contact response, while a penalty approach is introduced in the normal contact direction. The development of  the finite element (FE) formulation stemming  from its variational formalism is thoroughly derived and the model is validated in relation to challenging scenarios for standard (alternative) finite element procedures and analytical methods, such as the contact with multi-scale rough profiles. The present framework enables the comprehensive investigation of the system response due to the occurrence of   tangential tractions, which are at the origin of important phenomena such as wear and fretting fatigue, together with the analysis of the effects of coupling between normal and tangential contact tractions. This scenario is herein investigated in relation to challenging physical problems involving arbitrary loading histories. 
\end{abstract}


\end{frontmatter}

\section{Introduction}
The analysis of the tangential tractions arising during the frictional contacts between a rigid and an elastic body plays a central role in physics and engineering. Mathematical models based on different friction laws have been developed during the past years in parallel with the progresses in the field of contact mechanics~\cite{barber2020,vakis}. In general, due to the intrinsic characteristics of the frictional contact problem, non uniqueness issues might arise, depending on the coefficient of friction and the regularisation of the Coulomb law adopted~\cite{klarbring1990529,raous1,raous2,raous3}.

A rigorous analytical treatment of this problem permits to obtain a solution only in few selected cases, characterised by simple contacting geometries and constitutive behaviours.  For example, starting from the general class of problems addressed in the first paragraph, if the attention is restricted to $2D$, small displacements and the deformable material is elastic and isotropic, the formal problem results in a set of two integral equations, coupled by the first and second Dundurs' bimaterial constants $\alpha$ and $\beta$, equipped with a proper friction law~\cite{barber_contact}. If the attention is further restricted to instances in which the half-plane hypothesis can be invoked, the problem reduces to two  integral equations coupled by the $\beta$ parameter only. Under these assumptions, the effect of coupling causes a vertical pressure to induce also horizontal displacements, thus influencing the tangential tractions distribution which in its turn modifies the vertical displacements, having so an appreciable effect on the vertical tractions and the contact area distribution. The half-plane approximation is particularly relevant since, under this assumption, problems involving also the contact of two elastic bodies can be addressed, via the introduction of \emph{ad-hoc} elastic moduli.

Under these hypotheses, the pioneering investigations of Cattaneo~\cite{cattaneo} and Mindlin~\cite{mindlin} are particularly relevant. These studies seminally concern with the solution of the problem of two contacting elastic spheres. They independently showed that the tangential tractions distribution caused by friction could be expressed as a superposition of two normal tractions distributions. Later, still within the framework of uncoupled approaches, J{\"a}ger~\cite{jager1,jager2} and Ciavarella~\cite{ciava1,ciava2} extended Cattaneo-Mindlin results to the contact problem of half-planes with generic non compact boundaries under oblique loading. In~\cite{goodman1962}, Goodman introduced the hypothesis of semi-coupling, taking into account only the effect of normal tractions on horizontal displacements, but not vice-versa, allowing the independent treatment of the vertical pressures with the usual methods employed for frictionless problems, and then solving separately for tangential tractions. This framework has been exploited by Nowell \emph{et al.}~\cite{nowell} who obtained the analytical solution for the special case of a quadratic indenter acting upon an elastic half-space. In the same reference, they also developed a numerical scheme based on the inversion of the governing integral equations for solving the related fully coupled problem with generic tangential loading histories. Under the hypothesis of full coupling, Spence~\cite{spence} solved the problem of the indentation of an elastic half-space by an axisymmetric punch for a monotonic normal load. In particular, he solved the problem for a flat punch and a Hertzian indenter, and proved, thanks to a property of the governing equations, that the solution for a generic power law profile could be directly derived from that corresponding to the flat punch. As a consequence, under the same intensity of the normal load, the extension of the slip area is found to be the same for every indenter with a power-law profile. 

More recently,  numerical methods have experienced a considerable development in order to overcome the limitations of the analytical approaches. Without the aim of providing an exhausting literature review on this topic, credit to the main contributions in the development of this subject can be given to Klarbring~\cite{klarbring1} and Klarbring and Bj{\"orkman}~\cite{klarbring2}, Kalker~\cite{kalker1,kalker2} and Ahn and Barber~\cite{ahn}. Numerical schemes based on the Boundary Element Method (BEM) have been proposed and widely used for the quantitative evaluation of tangential contact problems under arbitrarily complex contacting geometries. This framework has proved to be very efficient from a computational point of view, since it only requires the discretisation of the domain boundary, without accounting for the bulk. 
Many implementations are available, see e.g Zhao \emph{et al.}~\cite{zhao}, Pohrt and Li~\cite{pohrt} and Willner~\cite{willner}. On the other hand, BEM-based algorithms are limited by the assumptions of linear elasticity, half-space approximation and homogeneity of the materials. Their extension to inhomogeneities~\cite{nelias}, non-linear interface constitutive response~\cite{Rey2017,Popov2017,Popov} or finite size geometries~\cite{finite,vollebregt2} are sometimes possible but yet exceptions.


The Finite Element Method (FEM) represents today a suitable modeling tool for overcoming the inherent shortcomings of BEM, since it can easily account, for example, for geometrical or material non-linearities and finite-size geometries~\cite{HPMR,PEI}. A shortcoming of FEM applied to both frictionless and frictional contact problems is represented by the treatment of complex contacting geometries, since a discretisation of both the interface and the bulk is requested by the method. In case of very irregular (rough) contact geometries, not only demanding computational resources must be employed for their adequate discretisation, but also convergence issues might arise in standard node-to-segment and segment-to-segment contact strategies, due to corner cases experienced by the contact search algorithms to detect the points in contact. Therefore, regularisation techniques can be employed~\cite{HPMR,wriggers}, but with the drawback of artificially smoothing out real high-frequency profile features relevant for the physics of contact. 

In this work, a relevant extension of the interface finite element with eMbedded Profile for Joint Roughness (MPJR) interface finite element introduced in~\cite{MPJR} is presented to model the fully coupled frictional contact problem between a deformable body and a rigid indenter, with a significant simplification of the contacting interface discretisation, obtained by considering an equivalent smooth interface, instead of explicitly modeling its geometry. The main idea striving the current developments concerns re-casting  the original geometry of the contacting profiles via a macroscopically smooth interface, which enables the generation of a straightforward meshing with linear finite elements, albeit preserving  the actual geometry  of the rough profile. This information is stored in terms of its analytical expression through the correction to the initial gap function with the exploitation of the  assumption of a rigid indenting profile. The mathematical formulation is detailed in Sec.~\ref{sec:2}. The advantage of the present approach over alternative meshing methodologies relies on the fact that the shape of the profile, including waviness or even roughness, does not need to be explicitly modeled by the finite element discretisation. In the proposed scheme, the exact interface geometry is embedded into a nominally flat interface finite element through an analytical and exact correction of the normal gap function. In such a way the finite element discretisation and boundary geometry can be taken as nominally smooth, with a significant advantage in terms of regularity of the boundary and simplified finite element discretisation, albeit preserving the exact profile shape in the computations. The full derivation of the proposed finite element is presented in Sec.~\ref{sec:3}, and has been implemented as a user finite element in the research finite element analysis program FEAP~\cite{FEAP}. The forthcoming contents  of the manuscript is arranged as follows. The validation of the proposed model is carried out in Sec.~\ref{sec:validation}, in reference to a benchmark problem, namely an analytical solution of a Hertzian contact problem obtained by assuming semi-coupling between the normal and the tangential directions, under different load scenarios. In Sec.~\ref{sec:5}, the method is exploited to solve cases for which no analytical solutions are available, and that are particularly challenging according to standard finite element techniques. In particular, coupled normal and tangential frictional contact problems with indenters having Weierstrass profiles as boundaries and increasing number of length scales, representative of multi-scale waviness, are investigated.

\section{Variational formulation for frictional contact problems with embedded roughness}\label{sec:2}
In the present section, the variational formulation which governs the normal and tangential contact of two bodies across a rough interface is presented. First of all, the strong differential formulation is recalled, with the equations describing the mechanics of the two bodies together, along with the Hertz-Signorini-Moreau conditions for normal contact and the Coulomb law for tangential contact at the interface. Subsequently, the weak form is derived, and the contact conditions are treated. A penalty approach is employed for the normal contact, while a regularised friction law is employed for tangential interactions, in line with finite element procedures~\cite{wriggers_book}.
The weak form provides the starting point for the derivation of the interface finite element, which is then presented in the following subsection. The current formulation is valid for $2$D domains, although the proposed framework might be extended to $3$D in a straightforward way.

\subsection{General framework} 
\begin{figure}[h!]
\centering
\includegraphics[width=.5\textwidth,angle=0]{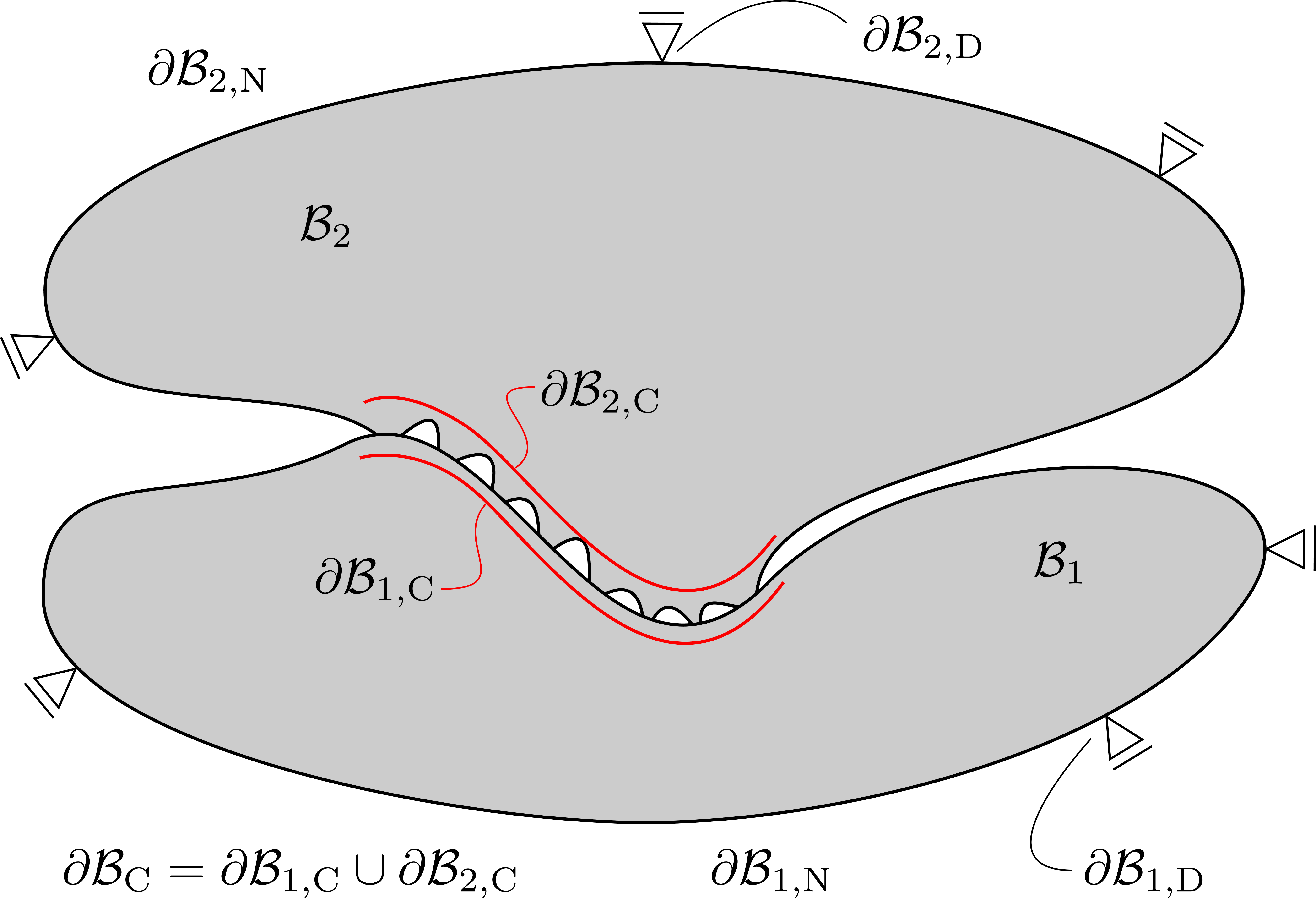}
\caption{domains $\Omega_i$ $(i=1,2)$, their Dirichlet $(\partial\Omega_i^D)$ and Neumann $(\partial\Omega_i^N)$ boundaries, and the contact interface $\partial\mathcal{B}_\mathrm{C}$.}\label{fig:domain}
\end{figure}

Let us assume that two deformable bodies define the domains $\mathcal{B}_i\in \mathbb{R}^2,\,i=[1,\,2]$ in the undeformed configuration defined by the Cartesian reference system $Oxy$. The boundary $\partial\mathcal{B}=\bigcup\limits_{i=1,2}\partial\mathcal{B}_i$ of the domain can be split into three distinct parts, Fig.~\ref{fig:domain}:
\begin{itemize}
\item[(i)]{a region where displacements are imposed, i.e. the Dirichlet boundary $\partial\mathcal{B}_i^\mathrm{D}$;}
\item[(ii)]{a region where tractions are imposed, i.e. the Neumann boundary $\partial\mathcal{B}_i^\mathrm{N}$;}
\item[(iii)]{an interface $\partial\mathcal{B}_\mathrm{C}$, common to the two bodies, where contact might take place, for which specific boundary conditions must be specified in order to model the stress and deformation fields generated by normal and tangential contact.}
\end{itemize}

As customary, a displacement field $\mathbf{u}_i(\mathbf{x},t) = [u_i(\mathbf{x},t),\,v_i(\mathbf{x},t)]^\mathsf{T}$, which maps the displacements of the points of $\mathcal{B}_i$ from the reference to the current configuration is assumed, identifying $u_i$ and  $v_i$ the displacements along $x$ and $y$ directions, respectively, and recalling  continuous and differentiable functions of the position vector $\mathbf{x}$ and time $t$. Further, a small deformation strain tensor is defined as the symmetric part of the deformation gradient, $\boldsymbol{\varepsilon}(\mathbf{u})_i : = \nabla^{S} \mathbf{u}_i$  (with $\nabla^{S}$ standing for the symmetric part of the gradient operator), which, in standard Voigt notation, reads $\boldsymbol{\varepsilon}_i = [\varepsilon_{xx},\,\varepsilon_{yy},\,\gamma_{xy}]_i^\mathsf{T}$.

At the interface $\partial\mathcal{B}_\mathrm{C}$, the configuration of the system is described by the relative displacement field, common for the two bodies, called \emph{gap field} across the interface, defined as $\mathbf{g}=\Delta\mathbf{u}$, which is the projection of the relative displacements of both bodies $(\mathbf{u}_2-\mathbf{u}_1)$ along the tangential and normal directions of the interface defined by the corresponding unit vectors $\mathbf{t}$ and $\mathbf{n}$, respectively. In components, the gap field vector reads $\mathbf{g} = [\gtt,\,\gnn]^\mathsf{T} = [\Delta\utt,\,\Delta\unn]^\mathsf{T}$.

\subsection{Definition of the equivalent contacting geometry}
\begin{figure}[h!]
\centering
{\includegraphics[width=.5\textwidth,angle=0]{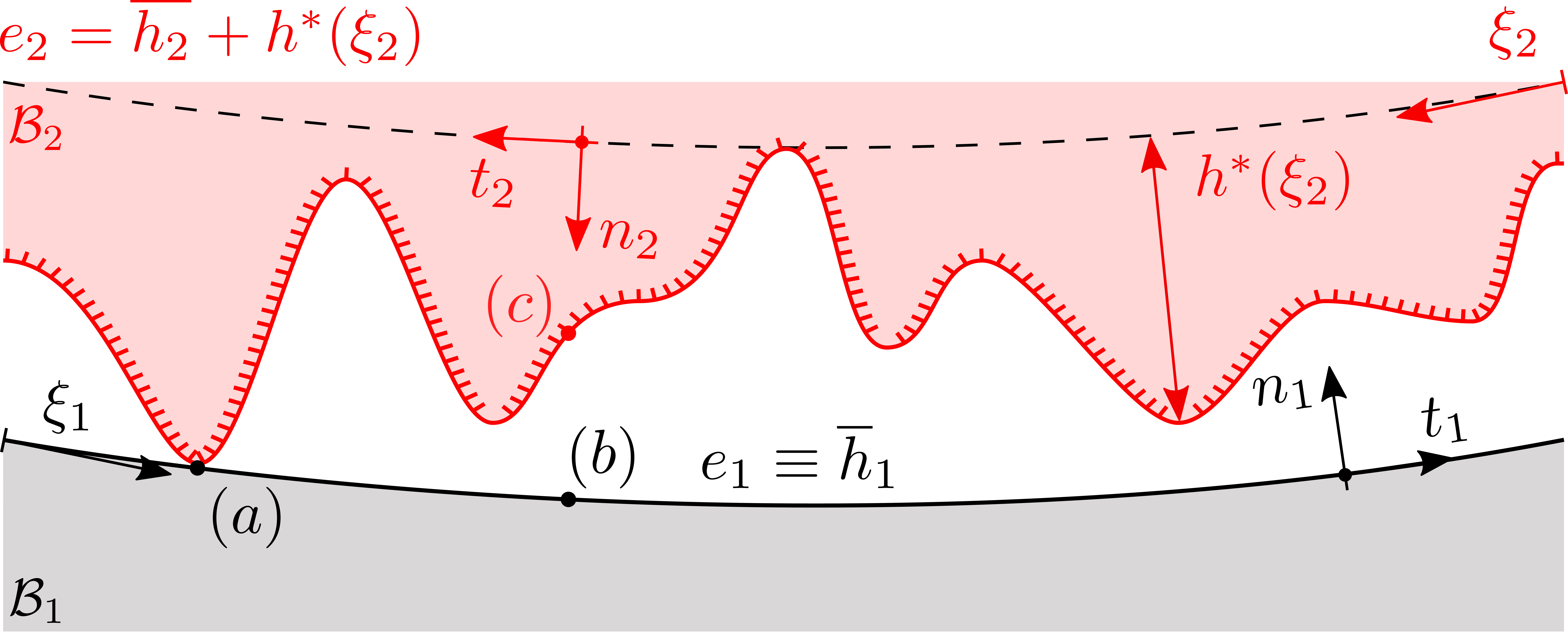}}
\caption{Composite topography of the interface embedding a rough profile}
\label{fig:comp1}
\end{figure}

The innovation of the proposed approach lies in the definition of the normal gap $\gnn$. In this framework, regardless of the actual topology, e.g. the one shown in Fig.~\ref{fig:comp1}, the contact interface is assumed to be \emph{nominally smooth}, with uniquely defined normal and tangential unit vectors, while its exact variation from planarity is analytically taken into account as a correction of the normal gap function.
First of all, a local coordinate system is introduced for both the boundaries of the two bodies, being $\xi_i = \xi_i(\mathbf{x})$ a curvilinear coordinate defining a point-wise correspondence with the coordinates of the same point in the global reference system. Tangential and normal unit vectors $\mathbf{t}_1(\xi_1)$ and $\mathbf{n}_1(\xi_1)$ can also be uniquely defined, the latter pointing outwards $\mathcal{B}_1$. If for every point of the contacting interface the indenter's profile local curvature radius is negligible with respect to the one of the deformable body, i.e. $R_1(\xi_1)\gg R_2(\xi_2)$, a \emph{smoother line} $\bar{h}_2(\xi_2)$ can be set, parallel to $\partial\mathcal{B}_{1,\mathrm{C}}$ and passing through the lowest point of elevation of $\partial\mathcal{B}_{2,\mathrm{C}}$, Fig.~\ref{fig:comp1}. A \emph{roughness function} $h^\ast(\xi_2)$ is then used to account for the actual profile elevation from the reference datum set in correspondence of $\bar{h}_2(\xi_2)$. Therefore, the actual profile elevations are described, in the local curvilinear reference system, by the \emph{elevation function} $e_2(\xi_2) = \bar{h}_2(\xi_2)+h^\ast(\xi_2)$. The smoothing line is parallel to $\partial\mathcal{B}_{1,\mathrm{C}}$, and a set of unit vectors $\mathbf{t}_2$ and $\mathbf{n}_2$ can now be defined, equal and opposite to their counterparts characterised by index $1$.
\begin{figure}[h!]
\centering
\includegraphics[width=.5\textwidth]{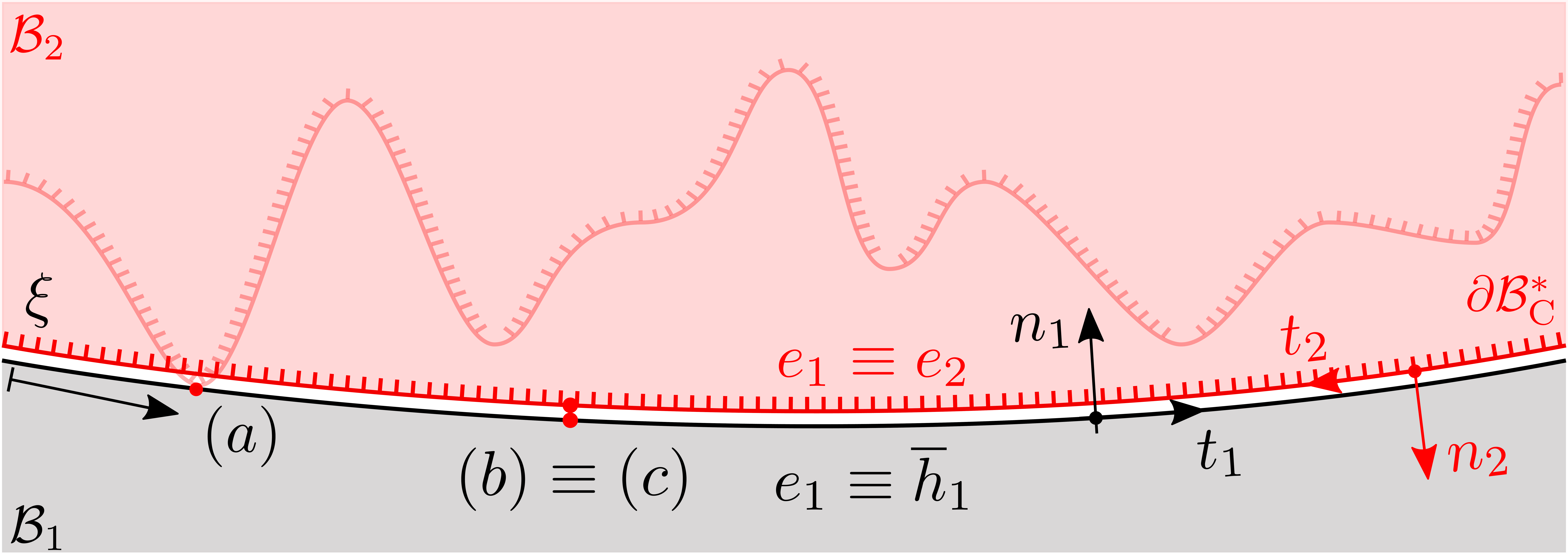}
\caption{Zero-thickness interface model defining the equivalent interface $\partial \mathcal{B}_\mathrm{C}^\ast$.}\label{fig:zt}
\end{figure}
With this transformation, a \emph{zero thickness interface} $\partial \mathcal{B}_\mathrm{C}^\ast$ is introduced. It is defined by two distinct lines which at the initial time $t=0$ are perfectly overlapping, and both defined by the function $e_1(\xi)$. A relative displacement field $\Delta \mathbf{u}$ can be defined based on the smooth interface of Fig.~\ref{fig:zt}, given by the relative displacements of matching points projected along the normal and tangential direction, e.g. points $(b)$ and $(c)$ of the same figure. The normal gap function $\Delta \unn$ defined in this way does not account for the actual complex geometry of the indenter. Its original morphology is stored in the function $h^\ast(\xi)$, which can be combined to $\Delta \unn$ with the aim of reconstructing the actual separation between the interfaces, such that it reads:
\begin{equation}\label{eq:interface}
\mathbf{g} = 
\left[
\begin{matrix}
\gtt\\
\gnn
\end{matrix}
\right] = 
\left[
\begin{matrix}
\Delta\utt\\
\Delta\unn+\max\limits_{\xi}[h^\ast(\xi)]-h^\ast(\xi)
\end{matrix}
\right].
\end{equation}
This operation can be performed thanks to the hypothesis of assuming the second body's material to be rigid, thus guaranteeing that its geometry is kept undeformed and can only undergo rigid body motion. If both the bodies are deformable, the error induced by the present approximation relies in two aspects: (i) the elastic contribution at the interface is locally affected by the geometry regularisation, which has however an average negligible effect for a nominally flat surface where valleys compensate peaks; (ii) the pointwise change in $h^\ast(\xi)$ due to local deformation is not accounted for unless an update in the geometry along with the deformation process is considered. Therefore the proposed methodology is expected to provide relatively good engineering approximations to the solution of the contact problem even in the most complex case where both materials are deformable.
The real portion of the boundary in contact is now determined by the value of the modified function $\gnn$, in accordance with the condition that contact occurs where $\gnn=0$. For example, this condition is recovered at $t=0$ assuming that the two profiles make contact only in correspondence of the highest asperity (point $(a)$ of Fig.~\ref{fig:zt}). At the present stage, a negative value for $\gnn$ is not admissible since it would imply material compenetration.

\subsection{Composite topography}
\begin{figure}[h!]
\centering
\includegraphics[width=.5\textwidth,angle=0]{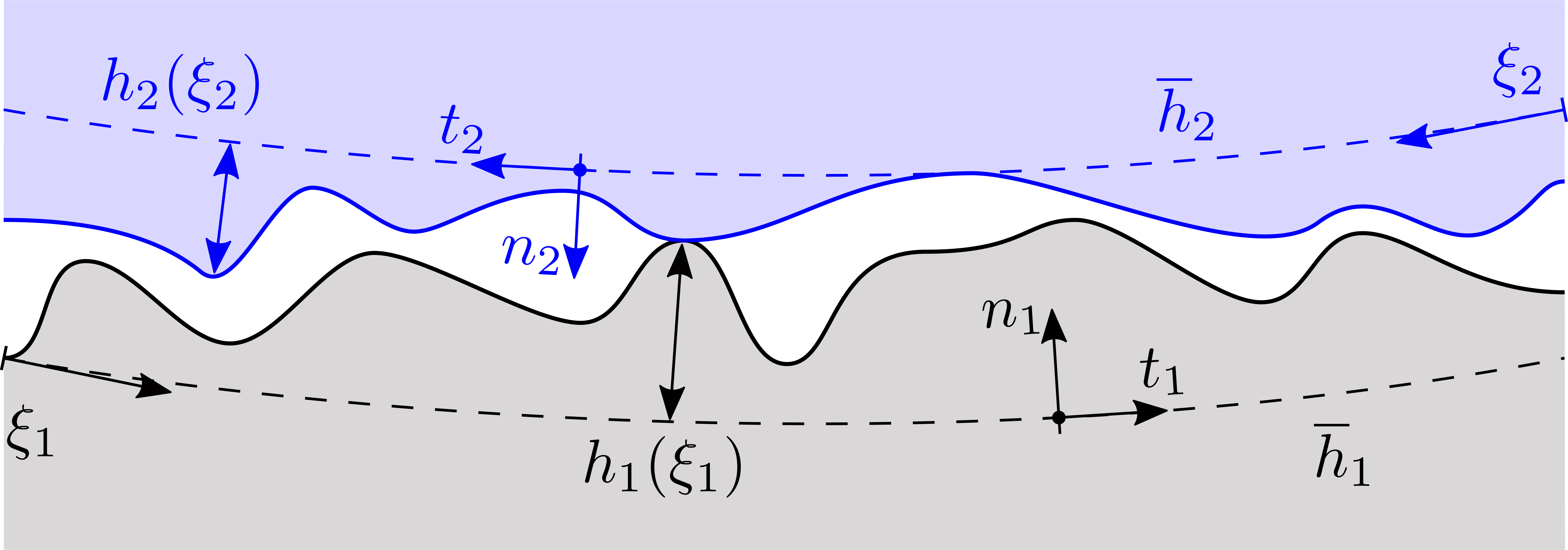}
\caption{Parametrisation of two rough profiles composing an interface $\Gamma_\mathrm{C}$.}\label{fig:comp}
\end{figure}

If the contacting bodies can be approximated as half-planes, e.g if the contact area is small compared to the bulks, the same formulation, i.e. rigid indenter over deformable body, can be interpreted as the solution of a contact problem involving two dissimilar elastic bodies, both potentially characterised by a geometrically complex interface. This problem can be re-cast in the already proposed framework by defining for the deformable body the composite elastic parameters $E^\ast$ and $\nu$, respectively given by: 
\begin{align}
\frac{1}{E^\ast} &= \frac{1-\nu_1^2}{E_1}+\frac{1-\nu_2^2}{E_2}, &
\nu &= \frac{1-2\beta}{2(1-\beta)},
\end{align}
in which $E_i$ and $\nu_i$ are the original Young's moduli and Poisson's ratios, and $\beta$ is the second Dundurs' constant.  A composite topography can be defined~\cite{ZBP04,ZBP07}, combining the geometries of the two contacting boundaries. Again two sets of curvilinear coordinates $\xi_i = \xi_i(\mathbf{x})$ can be defined, together with normal and tangential unit vectors $\mathbf{t}_i(\xi_i)$ and $\mathbf{n}_i(\xi_i)$, the latter pointing outwards $\mathcal{B}_i$. For each of the two profiles, a \emph{smoother line} $\bar{h}_i(\xi_i)$ is set, parallel to the average height of the original distribution and passing through its lowest point of elevation. Finally, the actual elevation of $\partial\mathcal{B}_1$ is flattened out, so that we have $e_1(\xi_1)\equiv\bar{h}_1(\xi_1)$, while the elevation of the indenter is defined as $e_2(\xi)=\bar{h}_2(\xi)+h^\ast(\xi)$, where now $h^\ast(\xi) = \max\limits_{\xi}[h_1(\xi)+h_2(\xi)]-[h_1(\xi)+h_2(\xi)]$, and $h_i(\xi_i)$ are the elevations measured from the respective datum, as shown in Fig.~\ref{fig:comp}. The common coordinate $\xi$, which is coincident with $\xi_1$ is here introduced, thus recovering the original case of Fig.~\ref{fig:comp1}.
\subsection{Governing equations and strong form}
The linear momentum balance equation for both $\mathcal{B}_1$ and $\mathcal{B}_2$, along with Dirichlet and Neumann boundary conditions on $\partial \mathcal{B}_i^\mathrm{D}$ and $\partial \mathcal{B}_i^\mathrm{N}$, can now be recalled for obtaining the strong form of equilibrium for the contacting bodies, enhanced by the conditions for normal contact on $\partial\mathcal{B}^\ast_\mathrm{C}$, and the classic Coulomb friction law for tangential contact, which determines its  further partition as $\partial\mathcal{B}^\ast_\mathrm{C} = \partial\mathcal{B}^\ast_\mathrm{C,st}\cup \partial\mathcal{B}^\ast_\mathrm{C,sl}$, where the subscripts \emph{st} and \emph{sl} stand, respectively for the \emph{stick} zone, where no tangential relative displacements between the two bodies occur, and the \emph{slip} zone, where irreversible tangential displacements, i.e. \emph{gross slip} or \emph{sliding}, take place~\cite{wriggers_book}:
\begin{subequations}\label{eq:strong}
\begin{align}
\nabla \cdot \bsigma_i &=\mathbf{0} \;\; \text{in}\,\mathcal{B}_i,\\
\bu_i &=\overline{\bu} \;\; \text{on}\, \partial\mathcal{B}_i^\mathrm{D},\\
\bsigma_i \cdot \mathbf{n} &= \mathbf{T}\;\; \text{on}\, \partial\mathcal{B}_i^\mathrm{N}, \\
\gnn &= 0,\, p_n<0\;\; \text{on}\, \partial\mathcal{B}^\ast_\mathrm{C}\\
\gtt &= 0,\, \lvert \ptt \rvert \le \mu\pnn\;\; \text{on}\, \partial\mathcal{B}^\ast_\mathrm{C,st}\\
\ptt &= -\mu\lvert \pnn \rvert\frac{\gttd}{\rvert \gttd\lvert},\, \lvert \ptt \rvert > \mu\pnn\;\; \text{on}\, \partial\mathcal{B}^\ast_\mathrm{C,sl}
\end{align}
\end{subequations}
where $\overline{\bu}$ denotes the imposed displacement and $\mathbf{T}$ the traction vector.

\subsection{Interface constitutive response}
Since not only stick occurs at the interface level, a single penalty approach cannot be used for both normal and tangential responses, thus requiring a distinction for the two different directions. In the normal one, a standard penalty approach has been used, which is characterised by a penalty parameter $\varepsilon_\mathrm{n}$  chosen high enough to reduce compenetration at the minimum. In the tangential direction, a regularisation of the Coulomb's law is introduced, as in~\cite{wriggers_book}. Given that, the two components of the interface traction vector $\mathbf{p} = [\ptt,\,\pnn]^\mathsf{T}$ are:
\begin{subequations}
\begin{align}
\pnn&=
\begin{cases}
\varepsilon_\mathrm{n} \gnn,\, \text{if}\; \gnn<0, \\
0,\, \text{if}\; \gnn\ge 0, \label{eq:pn}
\end{cases} &\\
\ptt &= \mu \lvert \pnn \rvert \tanh{\frac{\gttd}{\varepsilon_\mathrm{t}}}\label{eq:pt}.
\end{align}
\end{subequations}
Equation~\eqref{eq:pt} is one of the possible regularisations of Coulomb's friction law which enables a smooth transition from stick to slip condition. Correspondingly, if, on the one hand, it does not lead to a sharp transition from stick to slip domains, it has the great advantage of resolving the continuity issue of the Coulomb's law at the onset of slip. Finally, the presence of the regularisation variable $\varepsilon_\mathrm{t}$ allows tuning the curve in order to reproduce as closely as possible the real trend of the classic law.
It must be remarked that even with a regularised law, the introduction of friction at the interface level is still a source of non linearity for the problem, since tractions do depend on the displacements and therefore on the corresponding solution of the problem.

\subsection{Weak form}
The normal and tangential contact conditions on $\partial\mathcal{B}^\ast_C$, which modify the strong form expressed by Eq.~\eqref{eq:strong} with respect to the classic elastostatic set of equations, also determine the modification of the variational equality representing the weak form of the problem, resulting into the following variational inequality:
\begin{equation}\label{eq:weak_ine}
\sum_{\gamma=1}^2\Biggl[\int_{\mathcal{B}_\gamma}\bsigma_\gamma(\bu_\gamma) : \bepsilon_\gamma(\bv_\gamma)\,
\mathrm{d}\mathcal{V}_\gamma - \int_{\partial\mathcal{B}^\mathrm{N}_\gamma}\bar{\mathbf{t}}\cdot \bv_\gamma\,\mathrm{d}\mathcal{A}_\gamma\Biggl]\ge0,
\end{equation}
which is derived from the application of the principle of virtual work together with the application of the contact constraints. In Eq.~\eqref{eq:weak_ine}, $\bv_\gamma$ is the test function (virtual displacement field) for bodies $1$ and $2$, which fulfills the condition $\bv_\gamma=\mathbf{0}$ on $\partial\mathcal{B}_\gamma^\mathrm{D}$. Since body $2$ is rigid, its contribution to the integral could be omitted, however it has been taken into account for considering also the more general case of two dissimilar elastic bodies, easily recovering the limiting condition of body $2$ being rigid simply setting, in the practical application, a suitable ratio of their respective Young's moduli.
If the contact status is known, e.g if the contact problem is conformal or an active set strategy has been implemented for identifying the contact domain, then Eq.~\eqref{eq:weak_ine} can be recast in the form of an equality by adding the terms related to the normal and tangential contact constraints:     
\begin{equation}\label{eq:weak1}
\sum_{\gamma=1}^2\Biggl[\int_{\mathcal{B}_\gamma}\bsigma_\gamma(\bu_\gamma) : \bepsilon_\gamma(\bv_\gamma)\,
\mathrm{d} \mathcal{V}_\gamma - \int_{\partial\mathcal{B}^\mathrm{N}_\gamma}\bar{\mathbf{t}}\cdot \bv_\gamma\,\mathrm{d} \mathcal{A}_\gamma\Biggl]-C_\mathrm{n}-C_\mathrm{t}=0
\end{equation}
where, given the interface constitutive response chosen, the contributions related to normal contact, $C_\mathrm{n}$, and to friction, $C_\mathrm{t}$, read:
\begin{equation}\label{eq:contr}
\begin{aligned}
C_\mathrm{n}&=\int_{\partial\mathcal{B}^\ast_\mathrm{C}} \pnn(\bu) \delta g_\mathrm{n}(\bv)\,\mathrm{d}\mathcal{A},&
\\
C_\mathrm{t}&=\int_{\partial\mathcal{B}^\ast_\mathrm{C}} \ptt(\bu) \delta g_\mathrm{t}(\bv)\,\mathrm{d}\mathcal{A}.
\end{aligned}
\end{equation}
The displacement field $\bu_i$ solution of the weak form Eq.\eqref{eq:weak1} is such that it corresponds to the minimum of the energy for any choice of the test functions $\bv_i$.

\section{Interface finite element with eMbedded Profile for Joint Roughness (MPJR interface finite element) for frictional contact}\label{sec:3}
The numerical solution of the variational problem described by Eq.~\eqref{eq:weak1} in the framework of the finite element method requires the geometrical approximation of the two bulks, $\mathcal{B}_\gamma^\mathrm{h}$, and of the interface, $\partial\mathcal{B}^\ast_\mathrm{C}$, and their discretisation into finite elements:
\begin{subequations}
\begin{align}
\mathcal{B}_\gamma \approx \mathcal{B}_\gamma^\mathrm{h}&=\bigcup\limits_{\mathrm{e}=1}^{n_\Omega}\Omega_{\gamma,e}, & \\
\partial\mathcal{B}_\mathrm{C}^\ast \approx \partial\mathcal{B}_\mathrm{C,h}^\ast&=\bigcup\limits_{\mathrm{e}=1}^{n_\Gamma}\Gamma_e.
\end{align}
\end{subequations}
The bulk has been modeled using standard linear quadrilateral isoparametric finite elements~\cite{FEAP}, even though there is no restriction on the finite element topology, provided that it is consistent with that of the MPJR interface finite element used.

If the contact deformation is small, a conforming discretisation of the bulk and the interface is enough to guarantee the presence of pairs of nodes which are expected to come into contact. This assumption holds also in case of friction and relative tangential displacements, since slip will be infinitesimal too. Given that, a four nodes interface finite element $\Gamma_e$ can be introduced, as a special case of a collapsed $4$ nodes quadrilateral element. Its kinematics is borrowed from the formulation common in non-linear fracture mechanics for cohesive crack growth, see~\cite{OP99,PW11b,PW12,Reinoso2014,Paggi2015} and then specialised in the present case of frictional contact for the presence of complex surfaces.
\begin{figure}[h!]
\centering
\subfloat[][discretisation of the interface.\label{fig:}]
{\includegraphics[width=\textwidth,angle=0]{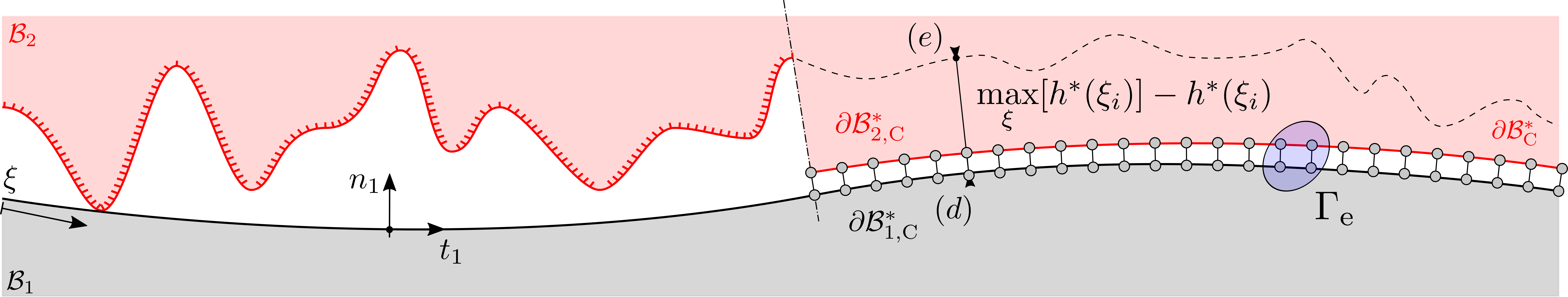}}\\
\subfloat[][Interface finite element.\label{fig:}]
{\includegraphics[width=.25\textwidth]{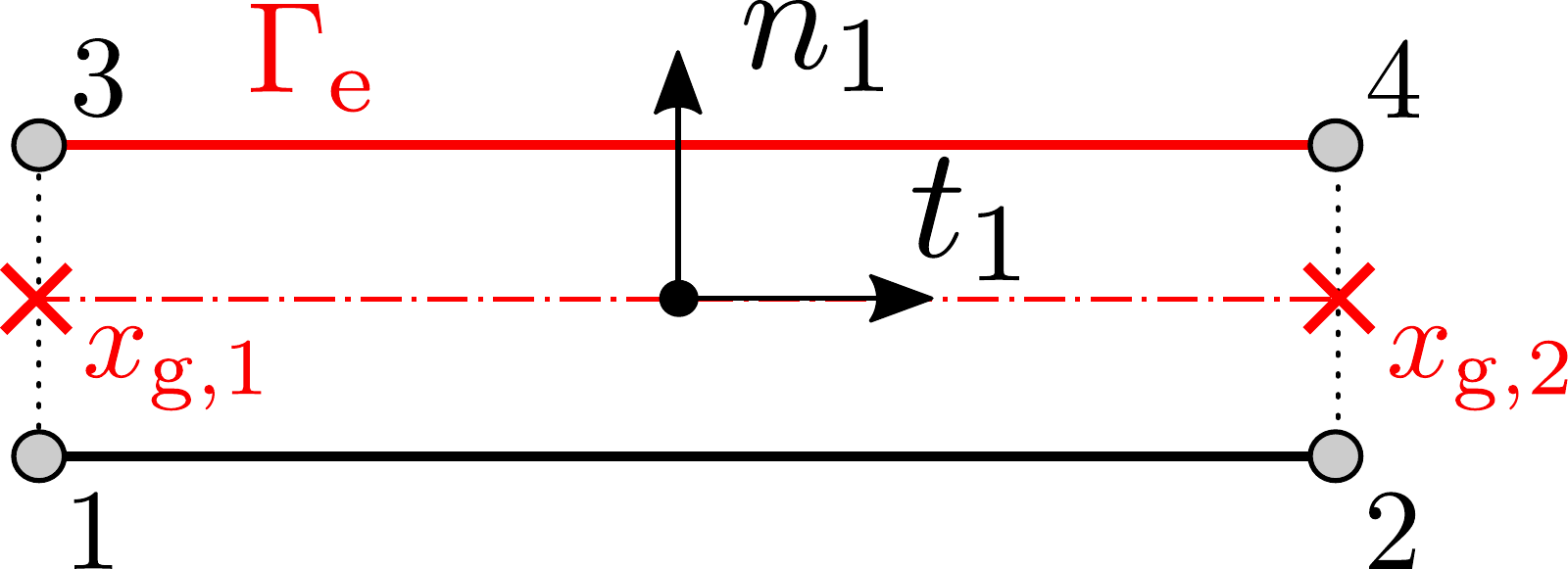}}
\caption{FEM approximation of the interface.}\label{fig:5}
\end{figure}
The interface element is defined by $4$ nodes, each pair belonging to the boundary of one of the two bodies: $1$ and $2$ to $\partial\mathcal{B}^\ast_{1,\mathrm{C}}$ and $3$ and $4$ to $\partial\mathcal{B}^\ast_{2,\mathrm{C}}$, see Fig.~\ref{fig:5}.
The contribution of the interface to the weak form is expressed by Eq.~\eqref{eq:contr}, whose geometrical approximation, together with the finite element discretisation, reads:
\begin{subequations}\label{weak-const}
\begin{align}
\int_{\partial\mathcal{B}^\ast_\mathrm{C}} \pnn(\bu) \delta g_\mathrm{n}(\bv)\,\mathrm{d}\mathcal{A}\approx
\int_{\partial\mathcal{B}^\ast_\mathrm{C,h}} \pnn(\bu) \delta g_\mathrm{n}(\bv)\,\mathrm{d}\mathcal{A}_\mathrm{h}&=
\bigcup\limits_{\mathrm{e}=1}^{n_\Gamma}\int_{\Gamma_e} \pnn(\hat{\bu}_\mathrm{e}) \delta g_\mathrm{n}(\hat{\bv}_\mathrm{e})\,\mathrm{d}\Gamma_e,&\\
\int_{\partial\mathcal{B}^\ast_\mathrm{C}} \ptt(\bu) \delta g_\mathrm{t}(\bv)\,\mathrm{d}\mathcal{A}\approx
\int_{\partial\mathcal{B}^\ast_\mathrm{C,h}} \ptt(\bu) \delta g_\mathrm{t}(\bv)\,\mathrm{d}\mathcal{A}_\mathrm{h}&=
\bigcup\limits_{\mathrm{e}=1}^{n_\Gamma}\int_{\Gamma_e} \ptt(\hat{\bu}_\mathrm{e}) \delta g_\mathrm{t}(\hat{\bv}_\mathrm{e})\,\mathrm{d}\Gamma_e.
\end{align}
\end{subequations}
In the equations above, the symbol $(\hat{\cdot})$ denotes, as customary, the difference between the exact displacement and its finite element approximation, while the subscript $e$ spans through the total number of elements $n_\Gamma$ employed for the interface discretisation.

The normal gap $\gnn$ and the tangential gap $\gtt$ are stored in the local gap vector $\mathbf{g}_\mathrm{l}=[\gtt,\,\gnn]^\mathsf{T}$. To evaluate them at every point of the interface element, the nodal displacement vector is introduced, defined as $\hat{\mathbf{u}}_\mathrm{e}=[u_1,v_1,\dots,u_4,v_4]^\mathsf{T}$, being $u_i$ and $v_i$ the respective horizontal and vertical displacements of the four nodes. A linear matrix operator $\mathbf{L}$ is deputed to evaluate the relative displacement $\Delta\hat{\mathbf{u}}_\mathrm{e}=[\Delta u_{\mathrm{e,t}},\,\Delta u_{\mathrm{e,n}}]^\mathsf{T}$ between nodes $1$ and $4$, and $2$ and $3$ respectively. A linear interpolation is then performed by the matrix multiplication with $\mathbf{N}(\xi)$, which collects the shape functions $N_1(\xi)$ and $N_2(\xi)$. The final step consists in the multiplication by the rotation matrix $\mathbf{Q}$, for moving from the global to the local reference system, centred in, and aligned to, the interface element. In formulae, it results:
\begin{equation}
\Delta\hat{\mathbf{u}}_\mathrm{e}=\mathbf{Q}\mathbf{N}\mathbf{L}\hat{\mathbf{u}}_\mathrm{e},
\end{equation}
and the matrix operators take the form:
\begin{align}
\mathbf{Q} &= 
\left[
\begin{matrix}
t_x & t_y\\
n_x   & n_y
\end{matrix}
\right],\notag & \\
\mathbf{N} &=
\left[
\begin{matrix}
N_1(\xi) &  0    &  N_2(\xi)  &  0 \\
    0 & N_1(\xi) &  0      &  N_2(\xi)
\end{matrix}
\right],\notag & \\
\mathbf{L} &=
\left[
\begin{matrix}
-1 &  0 &  0  &  0 &  0 & 0  &+1&  0\\
 0 & -1 &  0  &  0 &  0 & 0  & 0 &+1\\
 0 &  0 & -1  &  0 &+1 & 0  & 0 &  0\\
 0 &  0 &  0  & -1 &  0 &+1& 0 &  0
\end{matrix}
\right],\notag & \\
\hat{\mathbf{u}}_\mathrm{e} &=
\left[
\begin{matrix}
u_1&v_1&u_2&v_2&u_3&v_3&u_4&v_4
\end{matrix}
\right],\notag
\end{align}
being $t_x$, $t_y$ and $n_x$, $n_y$ the components of the unit vectors normal and perpendicular to the local reference system $\mathbf{t}_1$ and $\mathbf{n}_1$. The final tangential gap directly corresponds to $\Delta\hat{u}_\mathrm{e,t}$, the final normal gap is given by $\gnn = \Delta\hat{u}_\mathrm{e,n} + \max\limits_{\xi}[h^\ast(\xi)]-h^\ast(\xi)$, a correction which accounts for the actual complex surface of the rigid indenter. It has to be remarked that the normal gap is evaluated at every Gauss point, so two times for every interface element. In the discretisation of the contact zone, the number of interface element employed should be high enough in order to adequately sample and reproduce the elevation profile of the embedded contacting shapes. As a rule of thumb, in the case that low order elements are employed, at least $10$ elements should be employed for the accurate modeling of each asperity~\cite{yastrebov:2011}.

At this point it is possible to evaluate tractions $\pnn$ and $\ptt$ from Eqs.~\eqref{eq:pn} and~\eqref{eq:pt}, respectively.
The contribution of a single interface element to Eq.~\eqref{eq:weak1} can now be evaluated. Recalling the traction vector $\mathbf{p}=[\ptt,\,\pnn]^\mathsf{T}$, it is possible to condense Eq.~\eqref{eq:contr}, which for a single interface element read:
\begin{equation}
\delta\Pi_e = \int_{\Gamma_e}  \delta \mathbf{g}(\hat{\bv}_\mathrm{e})^\mathsf{T}\mathbf{p}(\hat{\bu}_\mathrm{e})\,\mathrm{d}\Gamma_e,
\end{equation}
where the variation of the local gap is given, in matrix notation, by:
\begin{equation}
\delta \mathbf{g}(\hat{\bv}_\mathrm{e}) = \frac{\partial \mathbf{g}}{\partial \hat{\bv}_\mathrm{e}}\delta \hat{\bv}_\mathrm{e} = \mathbf{Q}\mathbf{N}\mathbf{L} \delta \hat{\bv}_\mathrm{e}
\end{equation}
finally, the variation can be set to vanish, and the residual vector can be obtained: 
\begin{equation}
\delta\Pi_e = \delta \hat{\bv}_\mathrm{e}^\mathsf{T}\int_{\Gamma_e}\mathbf{L}^\mathsf{T}\mathbf{N}^\mathsf{T}\mathbf{Q}^\mathsf{T}\mathbf{p}(\hat{\bu}_e)\,\mathrm{d}\Gamma_e=0,
\end{equation}
which gives:
\begin{equation}
\mathbf{R}(\hat{\bu}_\mathrm{e}) = \int_{\Gamma_e}\mathbf{L}^\mathsf{T}\mathbf{N}^\mathsf{T}\mathbf{Q}^\mathsf{T}\mathbf{p}(\hat{\bu}_\mathrm{e})\,\mathrm{d}\Gamma_e = \mathbf{0}.\label{eq:res}
\end{equation}
A Newton-Cotes integration formula is exploited for evaluating the integral in $\mathbf{R}$, which requires the sampling of the integrand at points which coincide to the abscissae of nodes $1$ and $2$, obtaining:
\begin{equation}
\int_{\Gamma_e}\mathbf{L}^\mathsf{T}\mathbf{N}^\mathsf{T}\mathbf{Q}^\mathsf{T}\mathbf{p}(\hat{\bu}_\mathrm{e})\,\mathrm{d}\Gamma_e = \sum_{k=1}^2 w_k \mathbf{L}^\mathsf{T}\mathbf{N}_k^\mathsf{T}\mathbf{Q}^\mathsf{T}\mathbf{p}_k(\hat{\bu}_\mathrm{e})j_\mathrm{e}(\xi_k),
\end{equation}
being $w_k$ the weights and $j_\mathrm{e}(\xi)$ the determinant of the transformation mapping the change of coordinate from the global to the natural reference system. To light up the notation, hereinafter the vector of nodal displacement $\hat{\bu}_\mathrm{e}$ will be replaced by $\bu$.

Given the constitutive laws employed, Eq.~\eqref{eq:res} represents a set of non-linear transient equations. 
A full Newton Raphson iterative-incremental scheme is used to solve the system resulting from the discretisation of the rate problem relative to the bulk and the interface. The problem to be solved is therefore the following:
\begin{equation}
\mathbf{R}(\mathbf{u},\dot{\mathbf{u}}) = \mathbf{0},
\end{equation}
The residual vector is linearised introducing the tangent matrix, and the resulting set of linear equations reads:
\begin{equation}
\mathbf{S}^{(i)}\mathrm{d}\mathbf{u}^{(i)} = -\mathbf{R}^{(i)}
\end{equation}
where the superscript $i$ denotes the iteration inside a Newton Raphson loop and $\mathbf{S}$ is the tangent matrix defined as:
\begin{equation}
\mathbf{S} = - \frac{\partial \mathbf{R}}{\partial\mathbf{u}}-\frac{\partial \mathbf{R}}{\partial\dot{\mathbf{u}}}\frac{\partial \dot{\mathbf{u}}}{\partial\mathbf{u}} = c_1\mathbf{K}+c_2\mathbf{C},
\end{equation}
being $\mathbf{K}$ the stiffness matrix, $\mathbf{C}$ the damping matrix and $c_1$ and $c_2$ scalar coefficients that involve the time step $\Delta t$ and the parameters of the selected time integration scheme. For every cycle, the solution is updated:
\begin{equation}
\mathbf{u}^{(i+1)} =\mathbf{u}^{(i)}+\mathrm{d}\mathbf{u}^{(i)},
\end{equation}
until the convergence criterion $\mathrm{d}\mathbf{u}^{(i)}\cdot\mathbf{R}^{(i)}<\varepsilon$ is met. The stiffness and damping matrix resulting from the linearisation of the residual vector for a given iteration $i$ read:
\begin{subequations}
\begin{align}
\mathbf{K}^{(i)} &= \sum_{k=1}^2 w_k \mathbf{L}^\mathsf{T}\mathbf{N}_k^\mathsf{T}\mathbf{Q}^\mathsf{T}\mathbb{K}_k\,\mathbf{Q}\,\mathbf{N}_k\,\mathbf{L}\,j_\mathrm{e}(\xi_k), &\\
\mathbf{C}^{(i)} &= \sum_{k=1}^2 w_k \mathbf{L}^\mathsf{T}\mathbf{N}_k^\mathsf{T}\mathbf{Q}^\mathsf{T}\mathbb{C}_k\,\mathbf{Q}\,\mathbf{N}_k\,\mathbf{L}\,j_\mathrm{e}(\xi_k)
\end{align}
\end{subequations}
where $\mathbb{K}$ and $\mathbb{C}$ are the linearised interface constitutive matrices:
\begin{subequations}
\begin{align}
\mathbb{K} &=
\left[
\begin{matrix}
\dfrac{\partial \ptt}{\partial \gtt} &  \dfrac{\partial \ptt}{\partial \gnn}\\
\dfrac{\partial \pnn}{\partial \gtt} &  \dfrac{\partial \pnn}{\partial \gnn}
\end{matrix}
\right], &
\mathbb{C} &=
\left[
\begin{matrix}
\dfrac{\partial \ptt}{\partial \dot{\gtt}} &  \dfrac{\partial \ptt}{\partial \dot{\gnn}}\\
\dfrac{\partial \pnn}{\partial \dot{\gtt}} &  \dfrac{\partial \pnn}{\partial \dot{\gnn}}
\end{matrix}
\right].
\end{align}
\end{subequations}
$\mathbb{K}$ and $\mathbb{C}$ are derived by analytical differentiation if contact is detected, while on the other hand, every term of the matrices is set equal to zero if a positive $\gnn$, i.e. an open gap, is detected.

The application of the proposed approach offers advantages compared to standard FEM discretisation techniques. In particular, there is no need to discretise the interface geometry, since the deviation from planarity can be stored nodal-wise with a one-to-one correspondence between the original profile and the equivalent interface. This has a two-fold advantage: (i) convergence problems of contact search algorithms in case of spiky profiles can be avoided; (ii) the number of finite elements near the boundary can be significantly reduced.



\begin{figure}[t!]
\centering
\subfloat[][Standard FEM discretisation of a contact problem with\\ a wavy or rough boundary.\label{fig:gmsha}]
{\includegraphics[width=.5\textwidth]{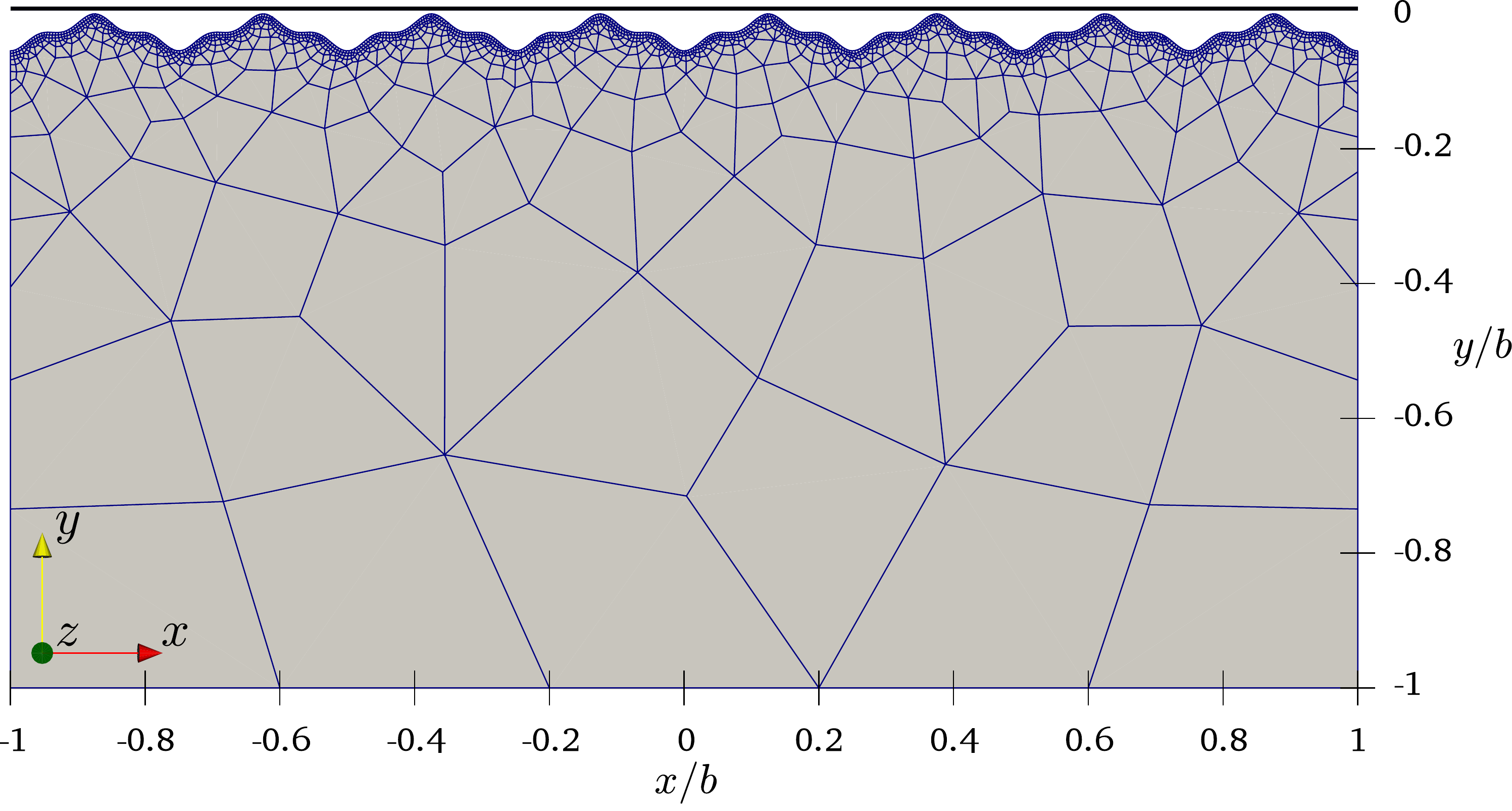}}
\subfloat[][MPJR FEM discretisation of the same problem of Fig.~6(a).\label{fig:WM_equiv}]
{\includegraphics[width=.5\textwidth]{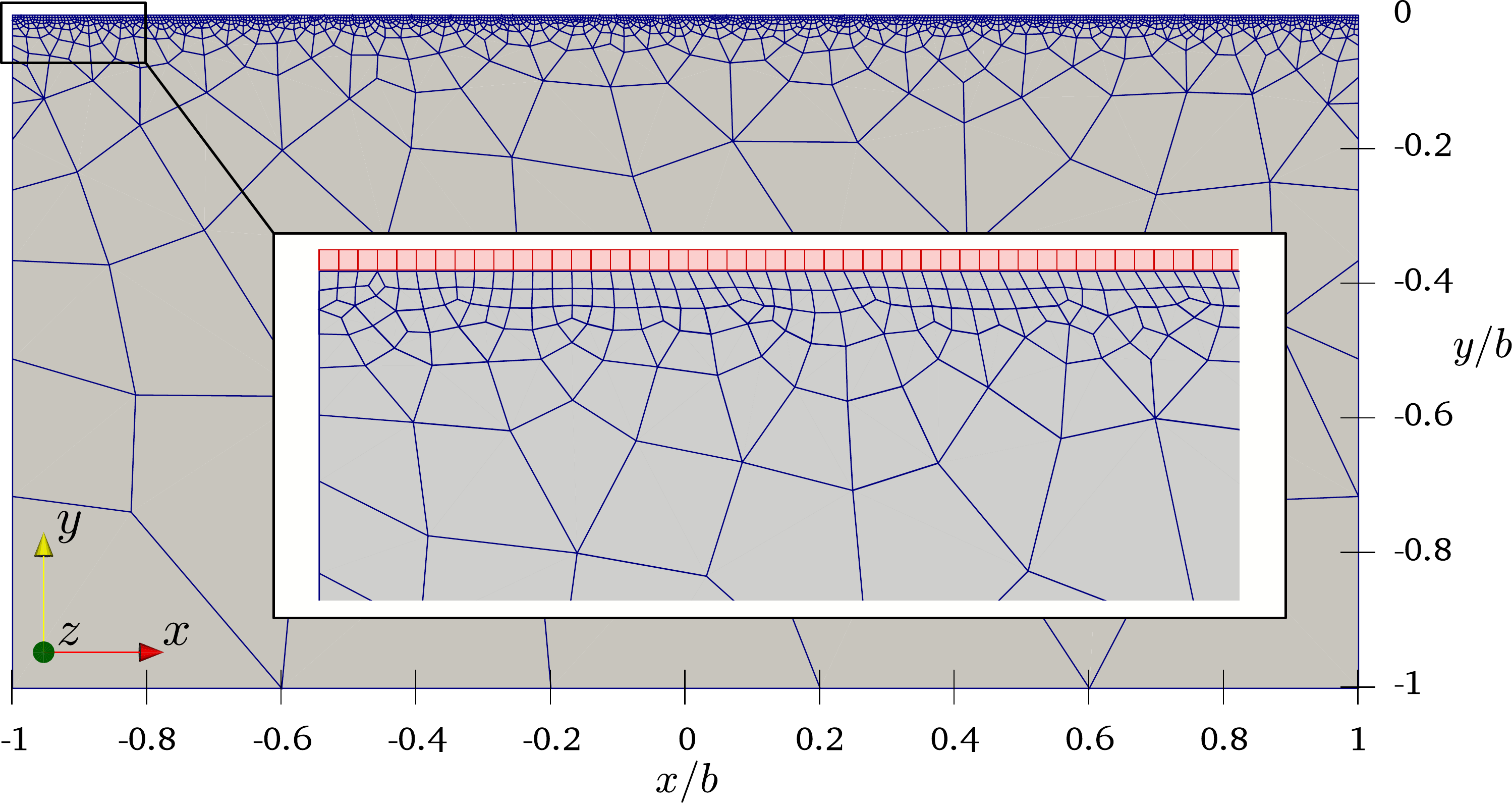}}
\caption{Comparison between the standard approach, Fig.~6(a) and the MPJR approach Fig.~6(b). The MPJR interface finite elements are shown in red in Fig.~6(b). In both cases a grading has been used such that $h_\mathrm{f}/b=4\times10^{-3}$ at the interface and $h_\mathrm{c}/b=4\times10^{-1}$ on the opposite side, being $b$ the height of the blocks and $h_\mathrm{f}$ and $h_\mathrm{c}$ the characteristic mesh sizes in correspondence of the interface and the lower side, respectively.}\label{fig:gmsh}
\end{figure}

As an illustrative example, let us consider the problem of an elastic block with a wavy boundary, pressed against a rigid flat plane, see Fig~\ref{fig:gmsha}. The boundary has been modeled as a Weierstrass profile made of two sinusoidal terms ($n_w=2$), see Sec.~\ref{sec:5} for details. The equivalent FEM discretisation resulting from the application of the MPJR approach is shown in Fig.~\ref{fig:WM_equiv}. In both cases, the same finite element mesh grading as been used, with a characteristic fine mesh size $h_\mathrm{f}/b=4\times10^{-3}$ in correspondence of the rough boundary, and a coarse mesh size $h_\mathrm{c}/b=4\times10^{-1}$ on the opposite side, being $b$ the overall height of the block. The finite element meshes have been generated using the open source mesh generator GMSH~\cite{geuzaine:2009}.

The use of the proposed approach is suitable to reduce the number of FEM nodes. At this regard, the reader is referred to the comparison presented in Tab.~\ref{tab:comparison}, where rougher profiles with an increasing number of sinusoidal terms have been considered. The gain is due to the fact that the MPJR discretisation does not need to exactly follow the wavy or rough profile as for the standard FEM discretisation, since it embeds the information about the rough topology as a correction to the gap function.

\begin{table}[h!]
    \begin{tabularx}{\linewidth}{@{}c X @{}}
    \includegraphics[width=0.4\linewidth,valign=c]{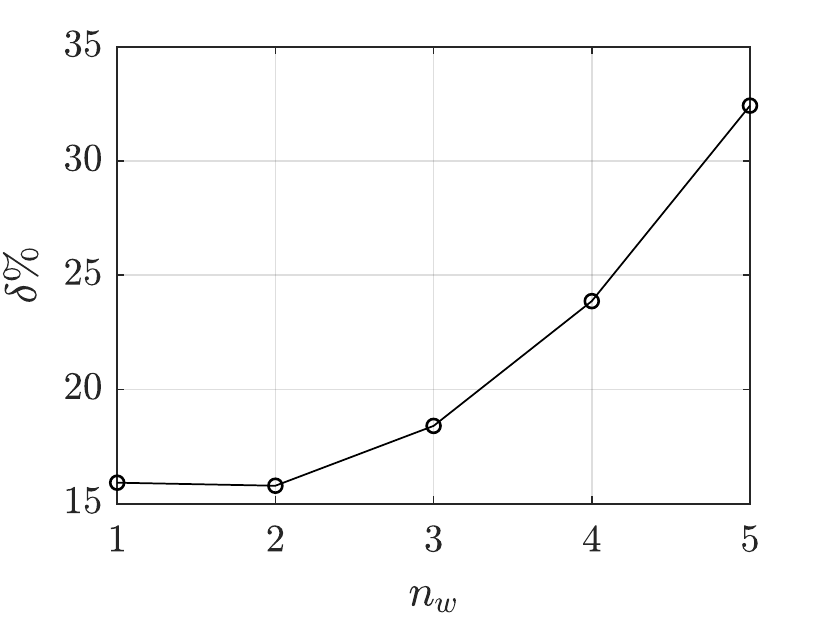}
    &   
\begin{tabular}{crr}
\toprule
{$n_w$}  & {method}  & {$n_\mathrm{nodes}$}  \\
\midrule\multirow{2}*{$1$} & \textsf{standard discr.}  & 1068  \\
                                                       & \textsf{MPJR}   & 898  \\
\midrule\multirow{2}*{$2$} &  \textsf{standard discr.} & 3136  \\
                                                       & \textsf{MPJR}  & 2641  \\
\midrule\multirow{2}*{$3$} &  \textsf{standard discr.} & 9552  \\
                                                       & \textsf{MPJR}   & 7794  \\
\midrule\multirow{2}*{$4$} &  \textsf{standard discr.} & 30621  \\
                                                       & \textsf{MPJR}   & 23312  \\
\midrule\multirow{2}*{$5$} &  \textsf{standard discr.} & 103198  \\
                                                       & \textsf{MPJR}   & 69730  \\
\bottomrule
\label{tab:comparison}
\end{tabular}
\end{tabularx}
\caption{Comparison in terms of number of finite element nodes ($n_{nodes}$) for a standard FEM discretisation, which explicitly models roughness, and for the proposed MPJR discretisation. The parameter $n_w$ denotes the number of terms of the Weierstrass profile, see Sec.~5. The side plot shows the relative percentage difference $\delta\%$ in the number of nodes required by the two different approaches, which grows together with the profile's complexity.}
\label{tab:comparison}
\end{table}

\section{Model validation}\label{sec:validation}
\subsection{Frictional Hertzian contact problem between a cylinder and a half-plane under a monotonic normal load}
\begin{figure}[t!]
\centering
\subfloat[][Actual geometry.\label{fig:6a}]
{\includegraphics[width=.3\textwidth,angle=0]{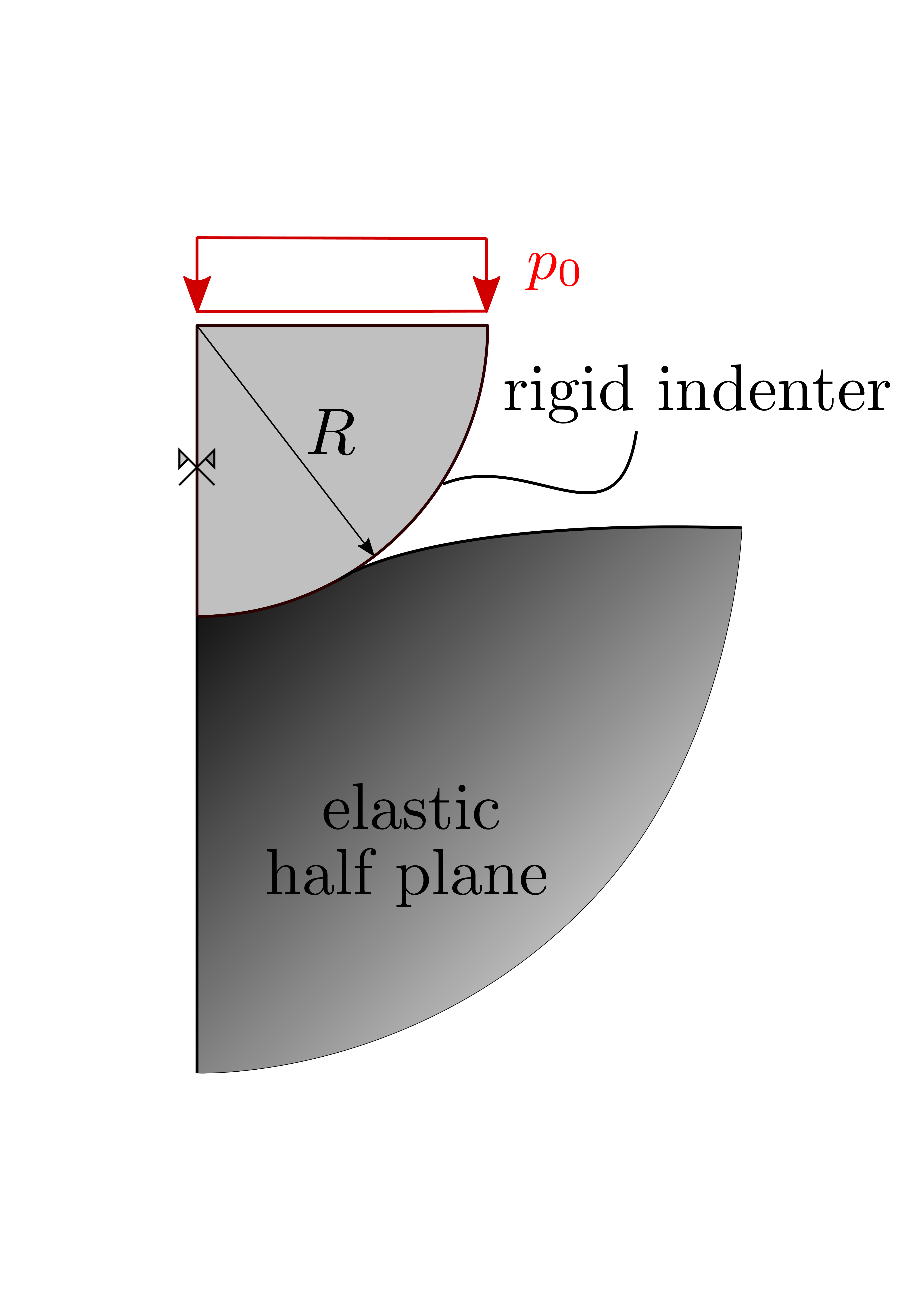}}\hspace{5mm}
\subfloat[][Finite element model.\label{fig:6b}]
{\includegraphics[width=.6\textwidth]{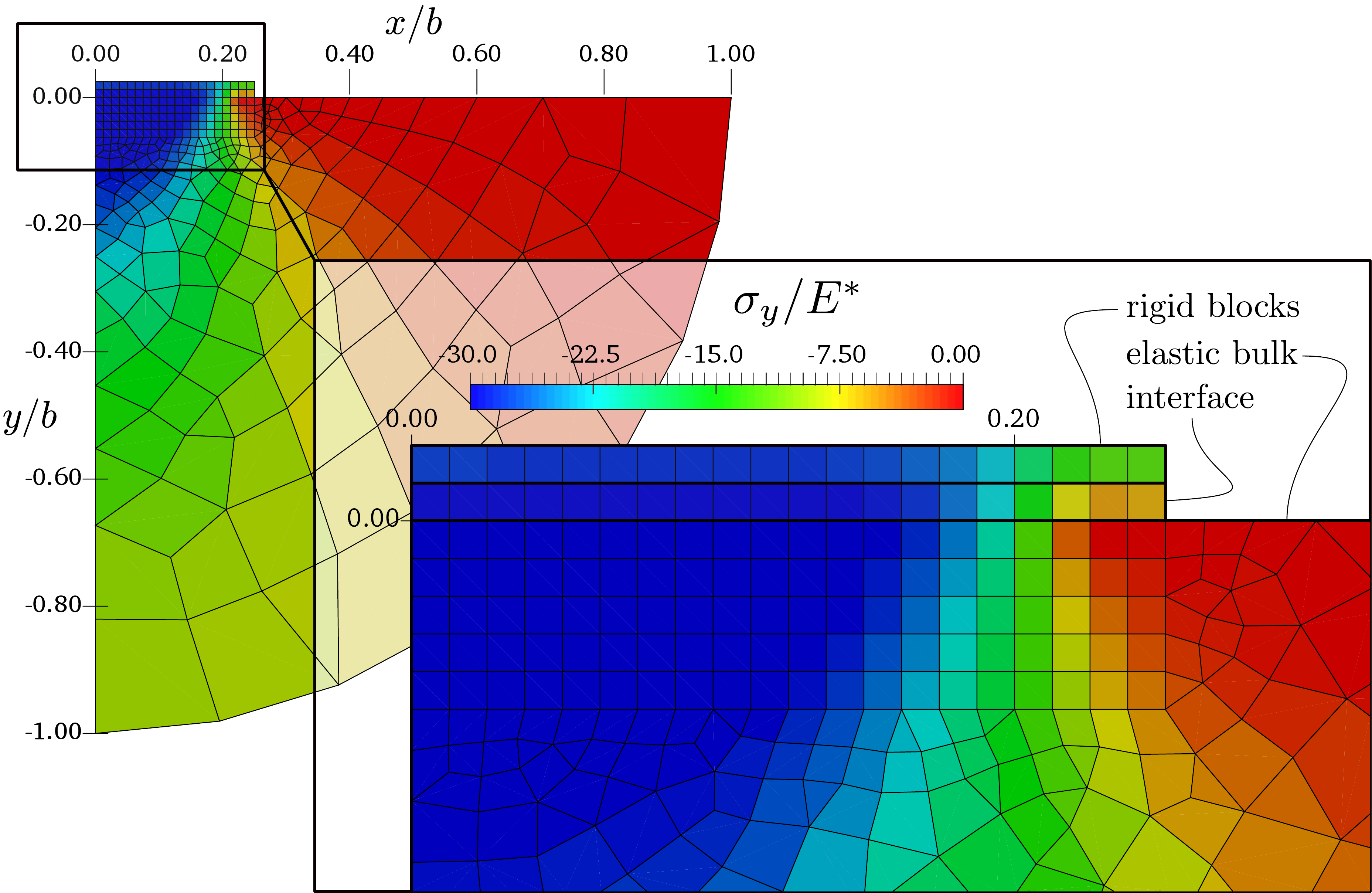}}
\caption{Actual geometry of the benchmark contact problem (a), and its finite element model set up with the present approach based on the MPJR interface finite element (b). In this figure, the model is characterised by a lower number of interface finite elements for clarity purposes, while the actual model employed for the validation has been discretised with a number of elements in accordance with Sec.~\ref{sec:FEM_validation}.}
\label{fig:hertz}
\end{figure}

A semi-coupled\footnote{Hereinafter, the label \emph{semi-coupled} will be referred to a system in which the effect of the tangential tractions over the vertical pressure is neglected, but not the opposite. This hypothesis has been introduced by Goodman~\cite{goodman1962}.} Hertz contact problem is used for validating the model, Fig.~\ref{fig:hertz}. This benchmark is found to be particularly suitable for testing the validity of the proposed implementation for two specific reasons. The first underlying motivation, already thoroughly explained in~\cite{MPJR}, is that in spite of its simplicity, the solution of this problem via a standard FEM discretisation of the curved geometry requires very refined meshes, specially at the edges of the contact strips. The second reason is that the present problem yields to a closed solution, thus providing an easy to implement and fast way of comparison and verification. Under the assumptions of purely normal monotonic load and neglecting the influence of tangential tractions $q_x(x)$ on the normal contact traction distribution $p_z(x)$, the solution can be written as~\cite{nowell}:
\begin{align}
q_x(x) &= \frac{\mu p_0}{K(c)}\Biggl(\sqrt{1-\frac{x^2}{a^2}}F(\theta,c)-\frac{x}{2b}\log\frac{1-\sqrt{1-x^2/b^2}}{1+\sqrt{1-x^2/b^2}}\Biggl), & b\le & \lvert x\rvert \le a\label{eq:semic}
\end{align}
where $p_0$ is the maximum value of the vertical tractions, $a$ and $b$ are the extension of the contact radius and slip zone respectively, $c=b/a$, $\sin(\theta)=x/b$ and $K(c)$ and $F(\theta,c)$ are the first complete and incomplete elliptic integral of the first kind, respectively. The extent of the slip zone can be evaluated using Spence's relation~\cite{spence}, which is valid for every initial contact gap defined by a power law relation, and is given by:
\begin{equation}
\mu K(c) = \beta K(\sqrt{1-c^2}),\label{eq:goodman}
\end{equation} 
where $\beta$ is the second Dundurs' constant. This parameter governs the level of coupling of the system: for $\beta=0$, the problem is uncoupled, while its maximum admissible value is $0.5$.
In the limiting case of a contact problem involving a rigid parabolic profile indenting an elastic half-plane, we have $\beta=(1-2\nu)/[2(1-\nu)]$, being $\nu$ the Poisson's ratio of the half-plane. A value of $0.29$ is herein chosen, which corresponds to $\nu=0.3$. While this value is kept constant, four different values for the friction coefficient have been used to investigate the accuracy of the model predictions. The comparison has been carried out in terms of normal and tangential tractions exploiting Eq.~\eqref{eq:semic}, comparing our fully coupled numerical approach with the semi-coupled analytical solution.

\subsection{FEM implementation and results}\label{sec:FEM_validation}
In the present framework, the actual shape of the contacting profile is embedded in the MPJR interface finite element. The profile elevation is given by its analytical form and evaluated at every Newton-Cotes integration point. Since one of the contacting profile, i.e. $\Gamma_1$, is flat, the topography exposed in Sec.~\ref{sec:2} simply reduces to the one of $\Gamma_2$: $e_2(\xi) = \bar{h}_2(\xi)-\xi^2/2R$ and we have a normal gap in the initial undeformed condition which is given by $\gnn = \Delta u_\mathrm{n} + \xi^2/2R$, being $R$ the radius of the cylinder. We analyse the problem under plane strain assumptions. The mesh is structured based on three different levels: 
\begin{itemize}
\item{the lower models $\mathcal{B}_1$, which in the present setting is a half-space, approximated by a circular sector clamped along the curved side and free to undergo vertical displacements in correspondence of the vertical side. An extension of its radius of $2R$ has been found to be enough for mimicking the elastic properties of a semi-infinite half-plane, under plane strain assumptions. A Young's modulus $E_1=100$ and a Poisson's ratio $\nu_1=0.3$ has been assigned to the standard quadrilateral linear finite elements employed for the bulk discretisation;}
\item{the interface $\partial\mathcal{B}_\mathrm{C}^\ast$ is modeled using a single layer of \emph{MPJR} elements, discretised using $n_\Gamma=100$ elements; The penalty stiffness has been set to $\varepsilon_\mathrm{n} = 10^2E_1/R$, that can be considered as sufficiently high in order to avoid material interpenetration.}
\item{finally, the geometry of the indenting cylinder which represents $\mathcal{B}_2$ can be replaced by a regular array of quadrilateral linear finite elements, with an assigned elastic modulus $E_2 = 10^3E_1$;
Neumann boundary conditions are applied as a uniform distribution of vertical pressure $p_0$ resulting in a unitary vertical force $P_z$.} 
\end{itemize}

\begin{figure}[t!]
\centering
\subfloat[][Normal and tangential contact tractions.\label{fig:low_c}]
{\includegraphics[width=.5\textwidth]{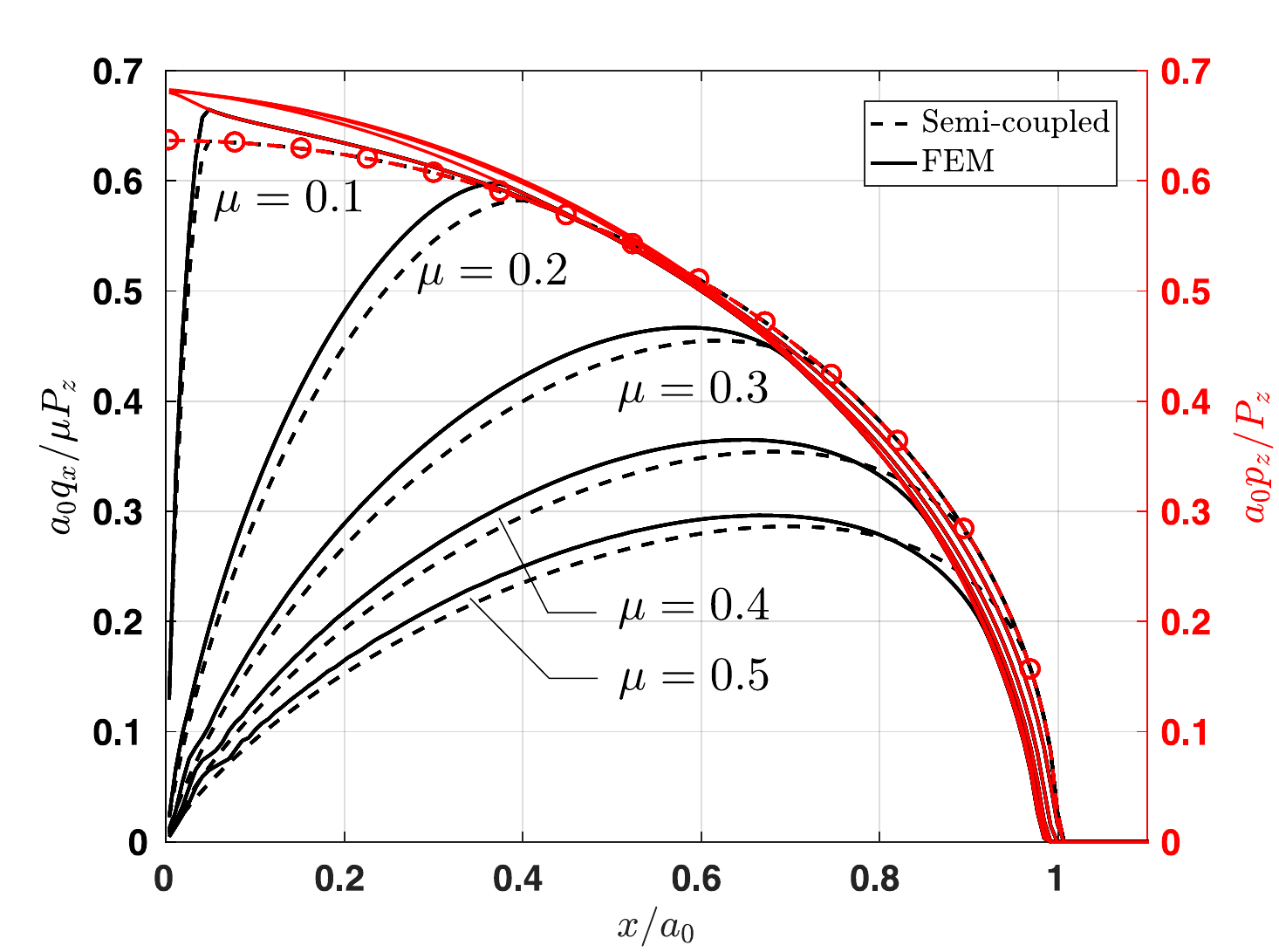}}
\subfloat[][Ratio between slip and stick contact strips width.\label{fig:goodman_curve}] 
{\includegraphics[width=.5\textwidth]{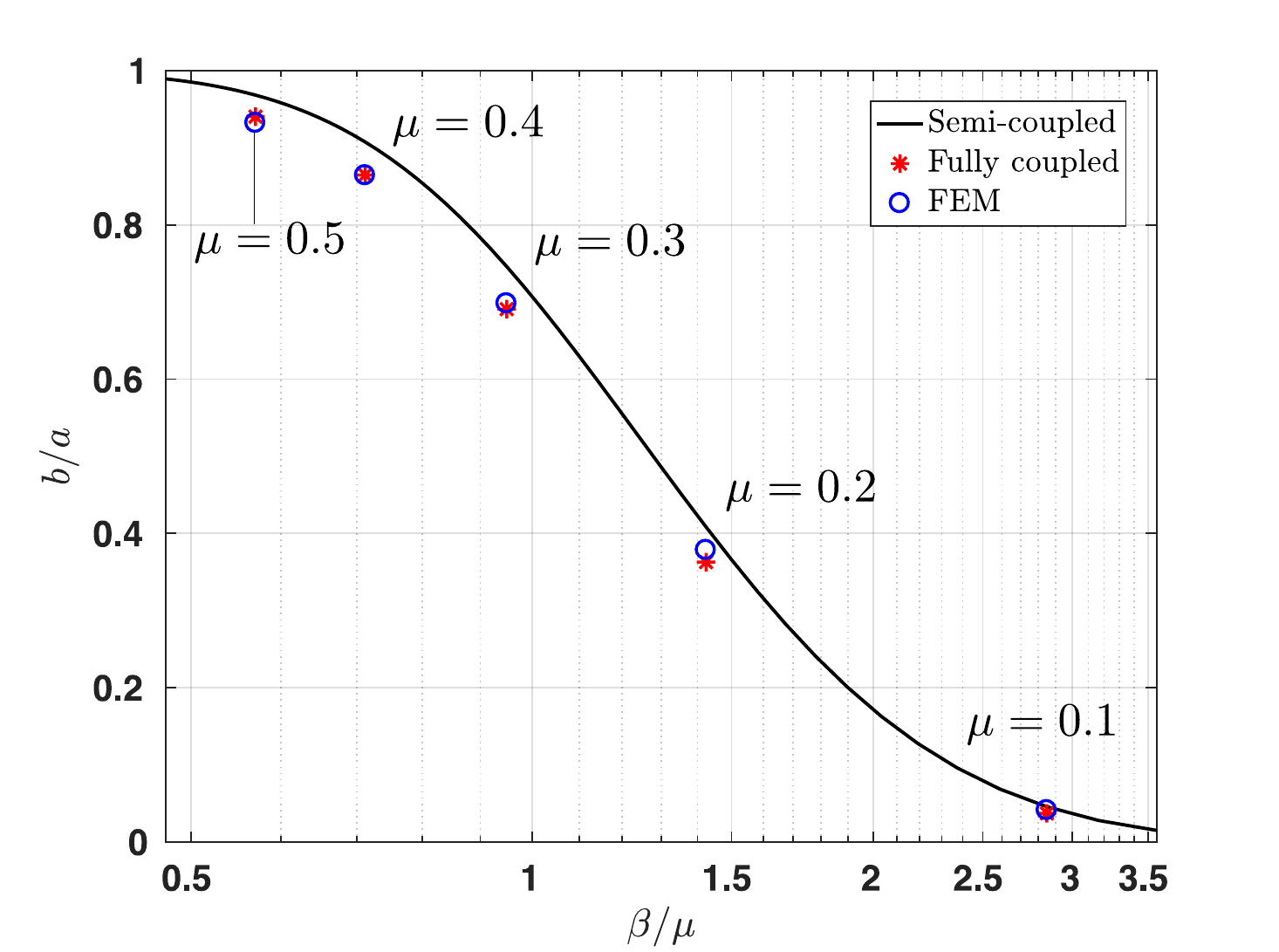}}
\caption{benchmark test with low values of coupling, $\beta=0.29.$}\label{fig:12}
\end{figure}

The results of the simulations are shown in Fig.~\ref{fig:low_c} for $\beta=0.29$, in terms of dimensionless vertical tractions $a_0p_z(x)/P_z$ (red curves) and dimensionless tangential tractions $a_0q_x(x)/\mu P_z$ (black curves), being $a_0$ the radius of contact related to the semi-coupled case and $P_z$ the vertical force applied. For both the analytical semi-coupled and the FEM model, five different values of the coefficient of friction, $\mu = [0.1,\,0.2,\,0.3,\,0.4,\,0.5]$ are applied. In the semi-coupled case, the distribution of vertical tractions is coincident regardless the coefficient of friction employed, and is highlighted by circle markers, while when coupling is considered, slight differences in the normal tractions can be observed. For what concerns the tangential tractions, an excellent accordance can be observed for all the values of the coefficient of friction employed. The coupling effect observed in the curve related to the numerical approach is in line with the theory, which predicts a stiffening effect that results into an increase of the maximum value of the vertical tractions, compensated by a slight decrease of the contact radius. The slight deviation between the two sets of $q_x(x)$ that can be observed in Fig.~\ref{fig:low_c} can be again ascribed to coupling. In fact, the FEM model predicts smaller values for the ratio $b/a$ with respect to the semi-coupled approach and the effect can be quantitatively analysed, with results shown in Fig.~\ref{fig:goodman_curve}. In this figure, the black solid curve represents the values of $b/a$ for the semi-coupled case, as expressed by Eq.~\eqref{eq:goodman}, corresponding to the dashed black curves of Fig.~\ref{fig:low_c}; the red stars are the values for the coupled case, evaluated using an asymptotic solution provided in~\cite{nowell} and the blue circles the outcomes of the simulation. A very good accordance is found between the slip/contact strip width ratio calculated in the fully coupled case and the one obtained by the numerical simulation; the deviation between the tangential tractions of Fig.~\ref{fig:low_c} can be justified in this way. The importance of the outcome of Fig.~\ref{fig:low_c} lies in the fact that, for some specific instances, care should be taken in neglecting coupling effects, since this could lead to underestimating the magnitude of tractions. As a final remark, it can be noticed that even if a regularised friction law has been used, an adequate choice of the stiffness parameter $\varepsilon_\mathrm{t}$ guarantees results that are very close to the solution based on the Coulomb friction law, which has been exploited by the semi-analytical model.

\subsection{Hertzian contact problem between a cylinder and a half-plane under constant normal loading and cyclic tangential loading}
\begin{figure}[t!]
\centering
\subfloat[][Load in the $t-\Delta$ space.\label{fig:loada}]
{\includegraphics[width=.5\textwidth]{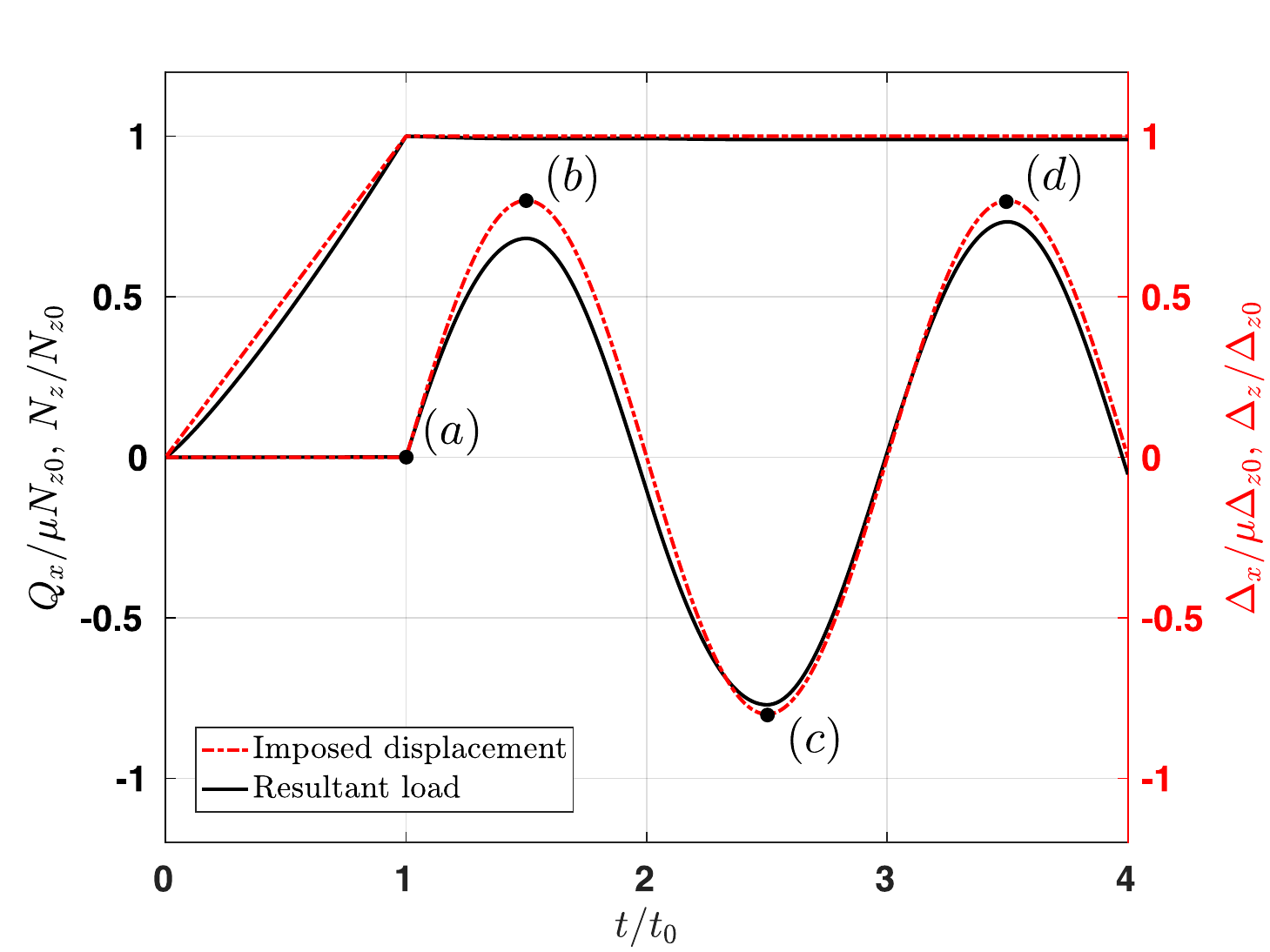}\label{fig:}}
\subfloat[][Load in the $\Delta_z-\Delta_x$ space.\label{fig:loadb}]
{\includegraphics[width=.5\textwidth]{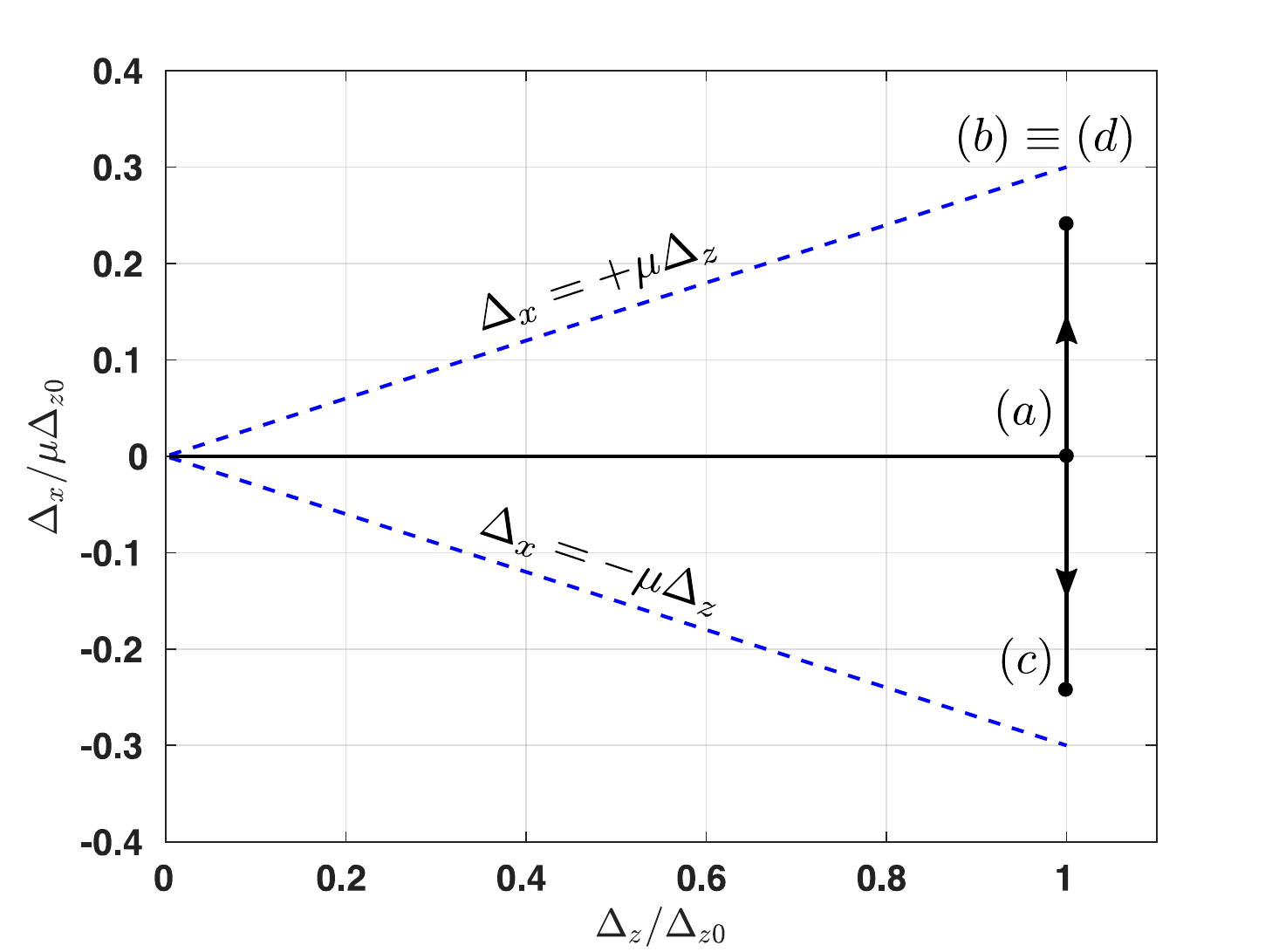}\label{fig:}}
\caption{tangential cycling load history.}\label{fig:load}
\end{figure}

\begin{figure}[h!]
\centering
\subfloat[][\label{fig:harm1a}]
{\includegraphics[width=.5\textwidth]{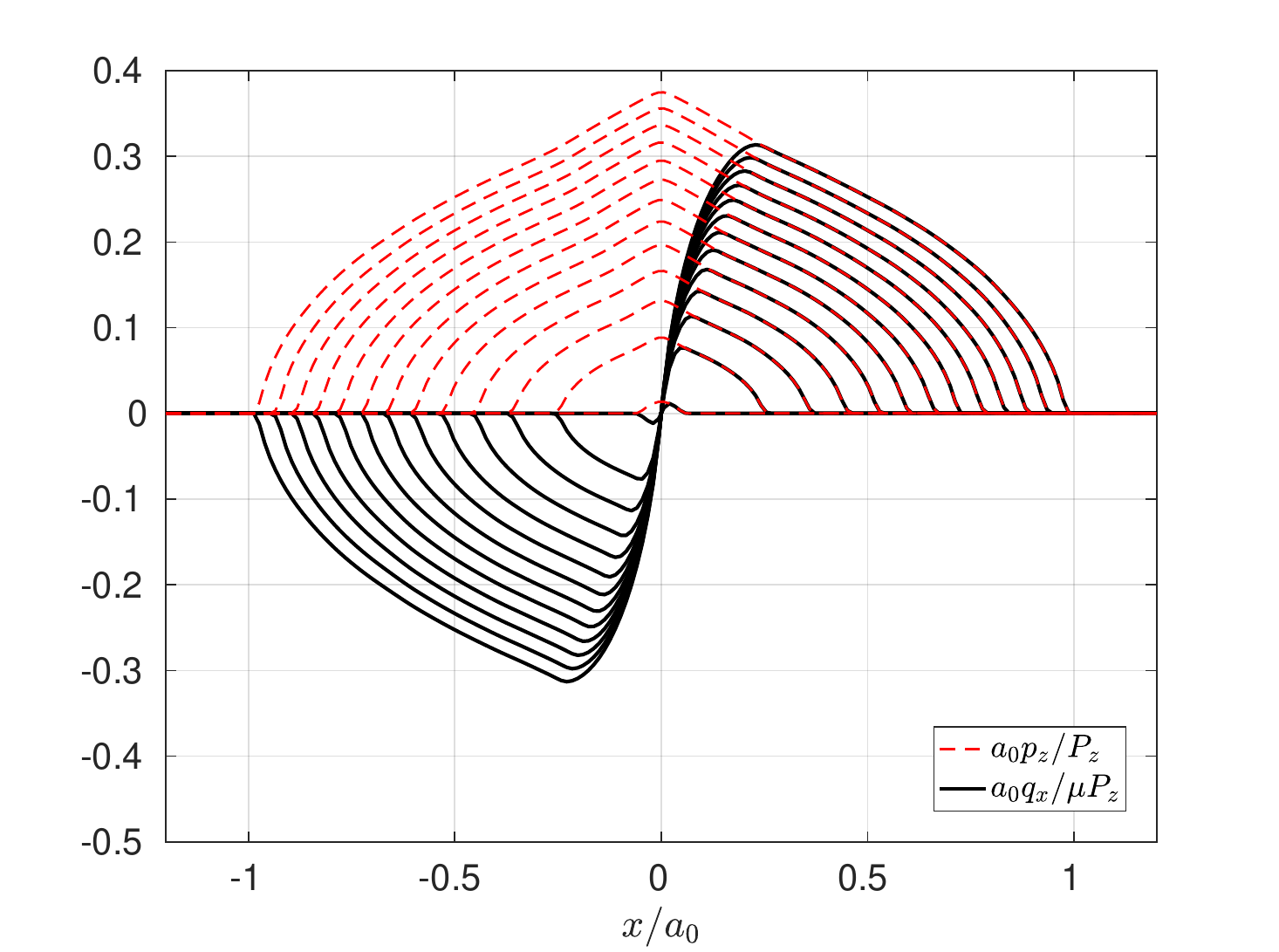}}
\subfloat[][\label{fig:harm1b}]
{\includegraphics[width=.5\textwidth]{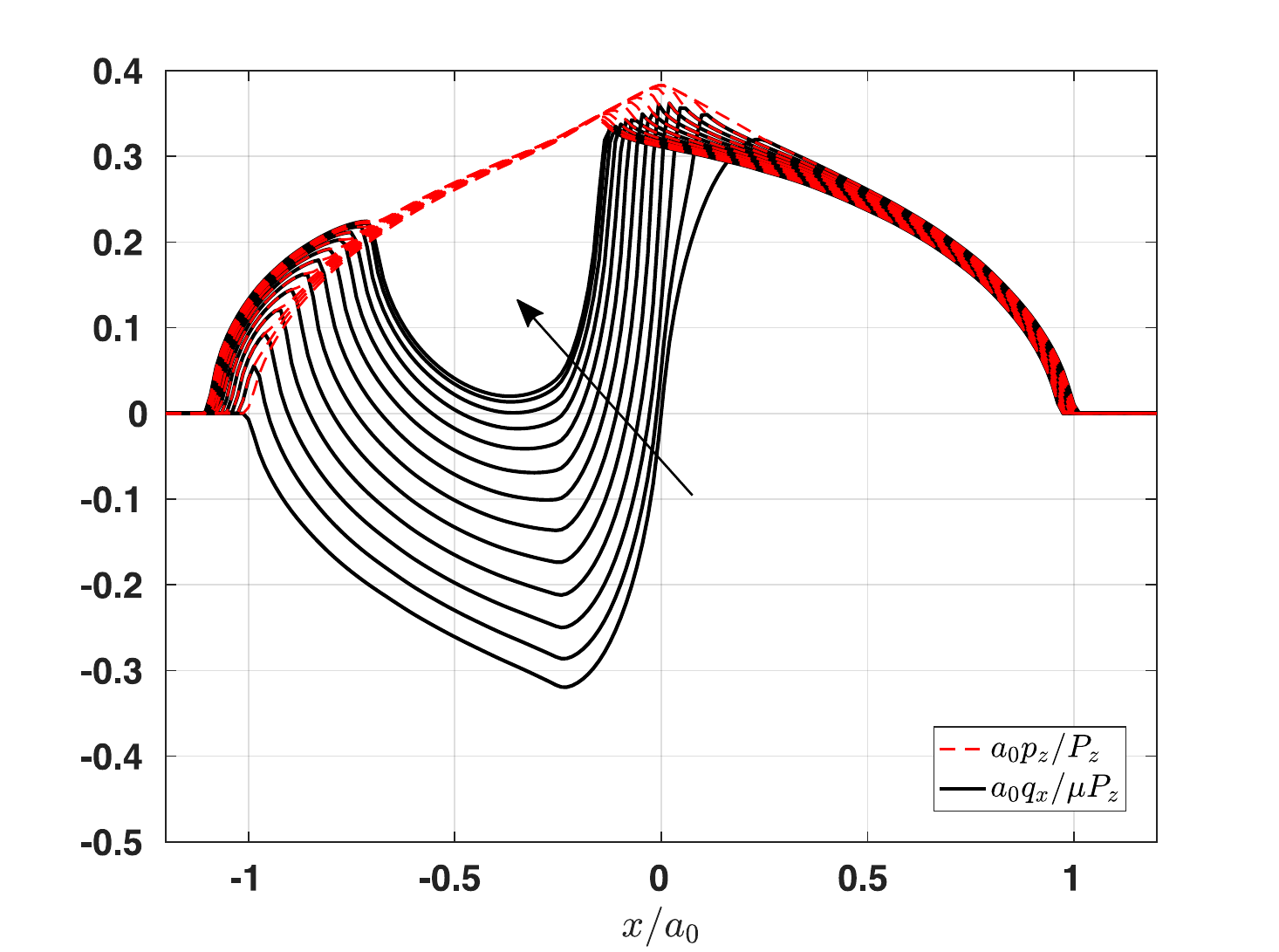}}\\
\subfloat[][\label{fig:harm1c}]
{\includegraphics[width=.5\textwidth]{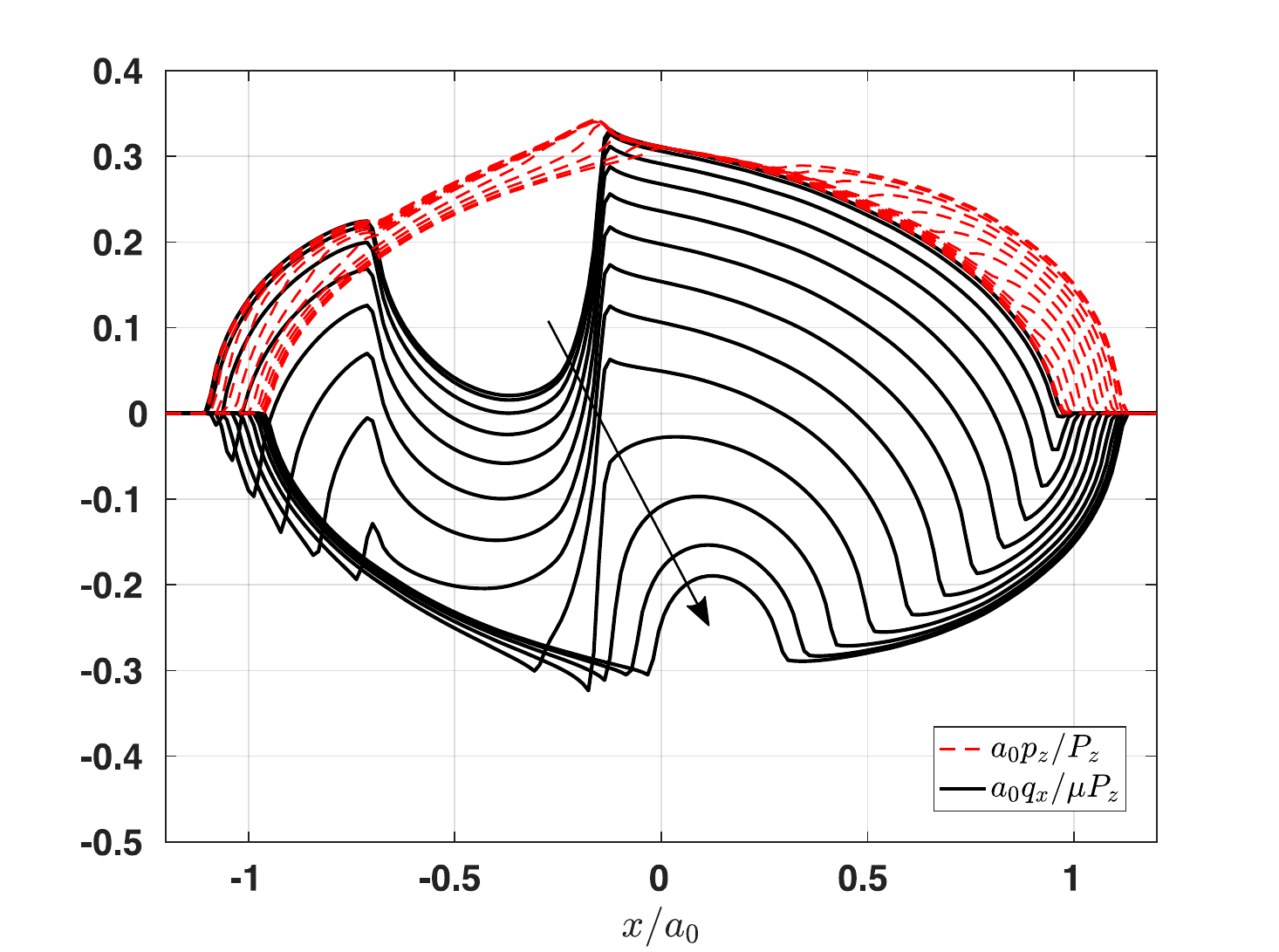}}
\subfloat[][\label{fig:harm1d}]
{\includegraphics[width=.5\textwidth]{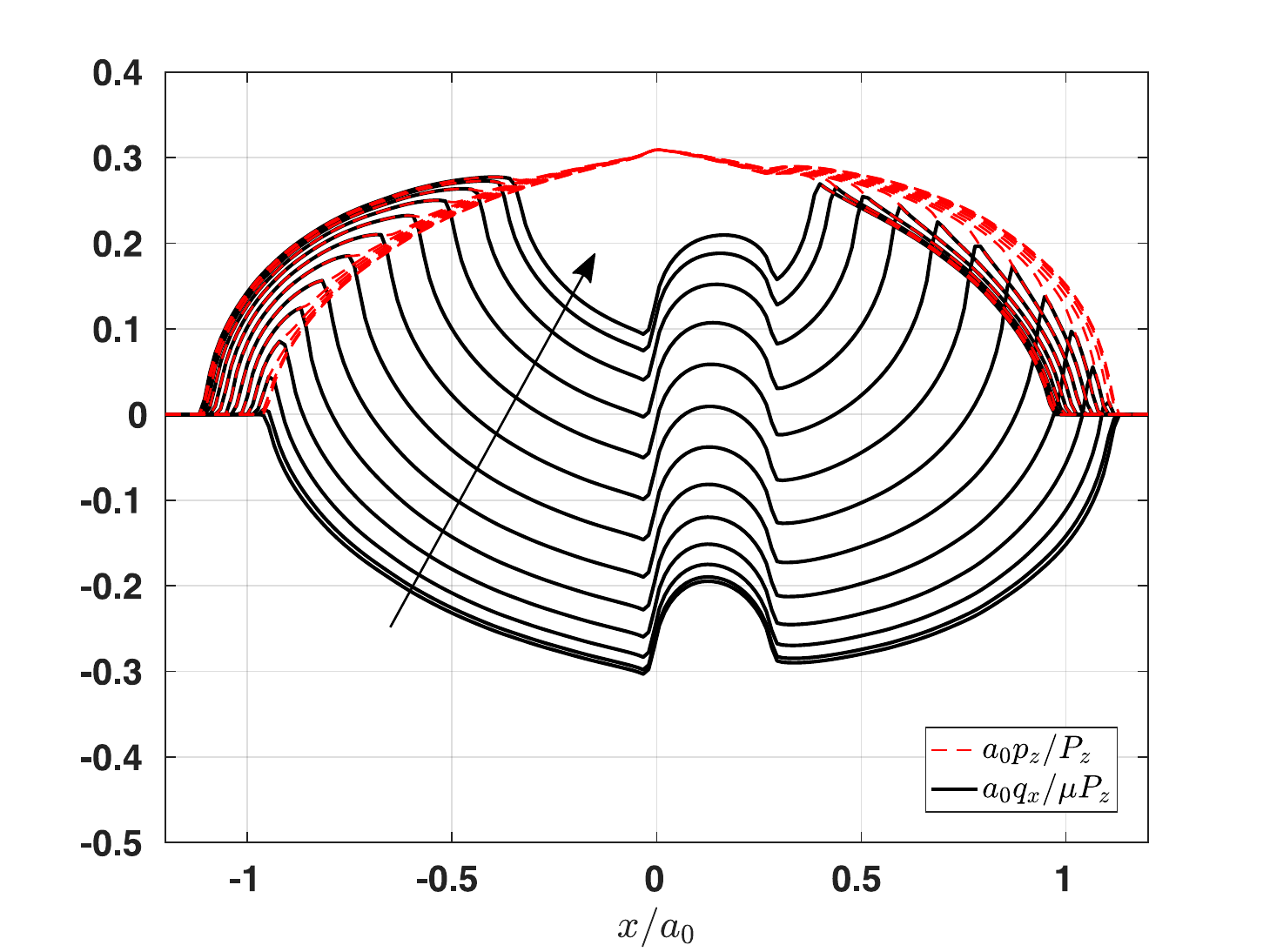}}\\
\caption{tangential cycling load history for high values of coupling, $\beta = 0.50$.}\label{fig:harm1}
\end{figure}

As compared to the previous test problem, now a downward displacement is imposed on the cylinder, starting from zero and linearly increasing up to a maximum value of $\Delta_{z,0}$. At this point, the vertical displacement is held constant and a tangential load is applied, in terms of a horizontal far field displacement which harmonically oscillates with amplitude $\Delta_{x,0} = 0.8\mu\Delta_{z,0}$. A normalised load history plot is shown in Fig.~\ref{fig:loada}, together with the corresponding values of total normal load $P_z$ and tangential load $Q_x$, evaluated as the resultant of the interface tractions. The imposed displacements are also plotted in the load space $\Delta_x-\Delta_z$, in which the black solid curve represents the variation of the tangential load with respect to the normal one, while the blue dashed curves represent the limit of gross sliding. In the present case the load path consists in a curve which is a straight line lying on the horizontal axis, from the origin to point $(a)$ and then becomes a collapsed ellipse with the major axis passing through the points $(b)$ and $(c)$, Fig.~\ref{fig:loadb}. The results in term of traction distributions are shown in Fig.~\ref{fig:harm1}. In Fig.~\ref{fig:harm1a}, the normal load is linearly increased from zero to its maximum value, and a self-similar symmetric central stick area encompassed by two regions of forward slip (positive tangential tractions) and backward slip (negative tangential tractions) develops. Then, Fig.~\ref{fig:harm1b}, the tangential load is applied (point $(a)$ of Fig.~\ref{fig:loada} and~\ref{fig:loadb}), which results in:
\begin{itemize}
\item[\emph{(i)}]{an increase in the tangential tractions at the leading edge;}
\item[\emph{(ii)}]{a reduction of the tangential tractions at the trailing edge;}
\item[\emph{(iii)}]{an instantaneous violation of the Coulomb law, which leads to a sudden change from backward slip to stick at the trailing edge, from which tractions start increasing again, but with opposite sign. At the same time, the stick zone shrinks, reaching its minimum at the point $(b)$ of the loading history.}
\end{itemize}

\begin{figure}[h!]
\centering
\subfloat[][\label{fig:harm2a}]
{\includegraphics[width=.5\textwidth]{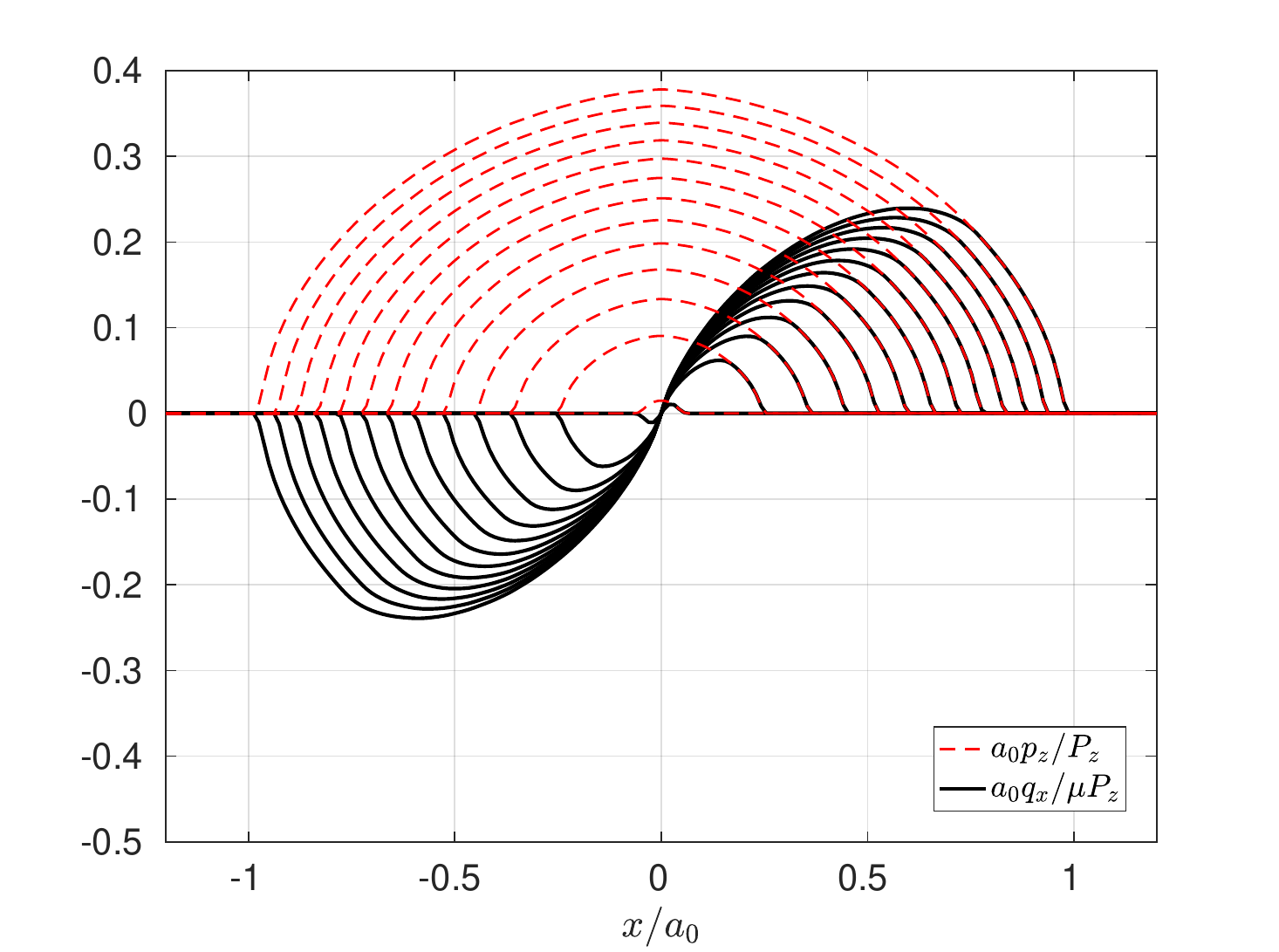}}
\subfloat[][\label{fig:harm2b}]
{\includegraphics[width=.5\textwidth]{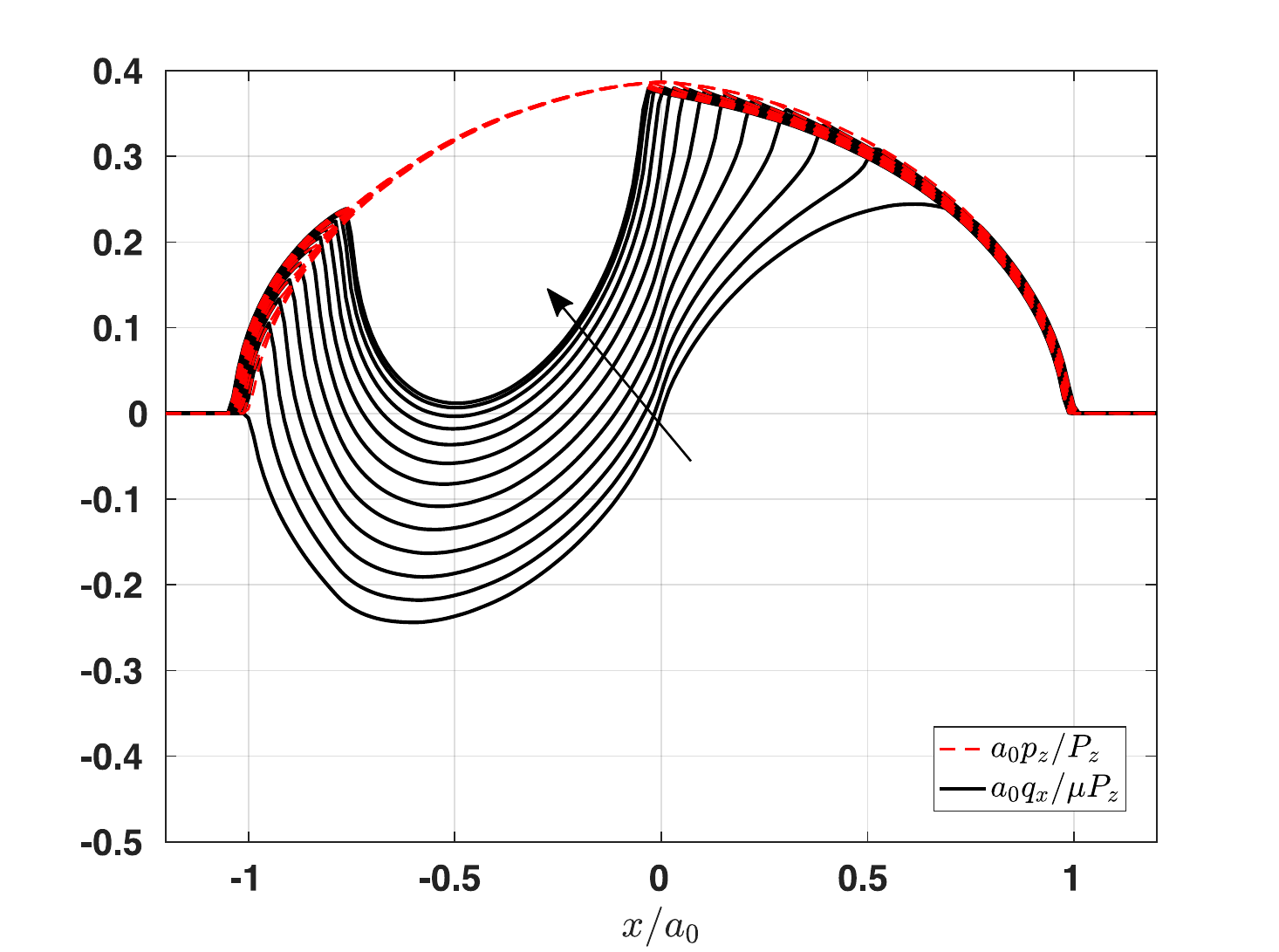}}\\
\subfloat[][\label{fig:harm2c}]
{\includegraphics[width=.5\textwidth]{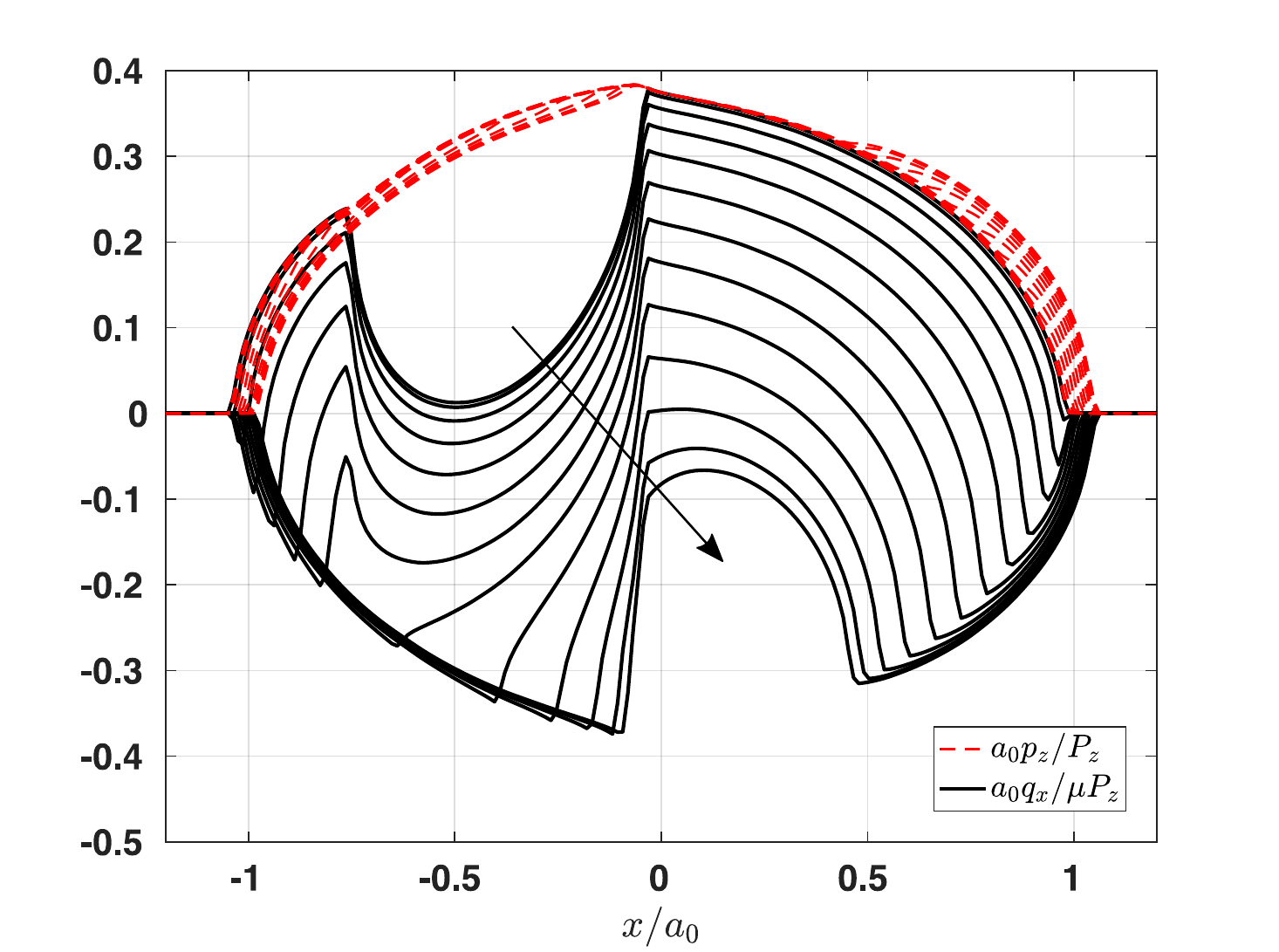}}
\subfloat[][\label{fig:harm2d}]
{\includegraphics[width=.5\textwidth]{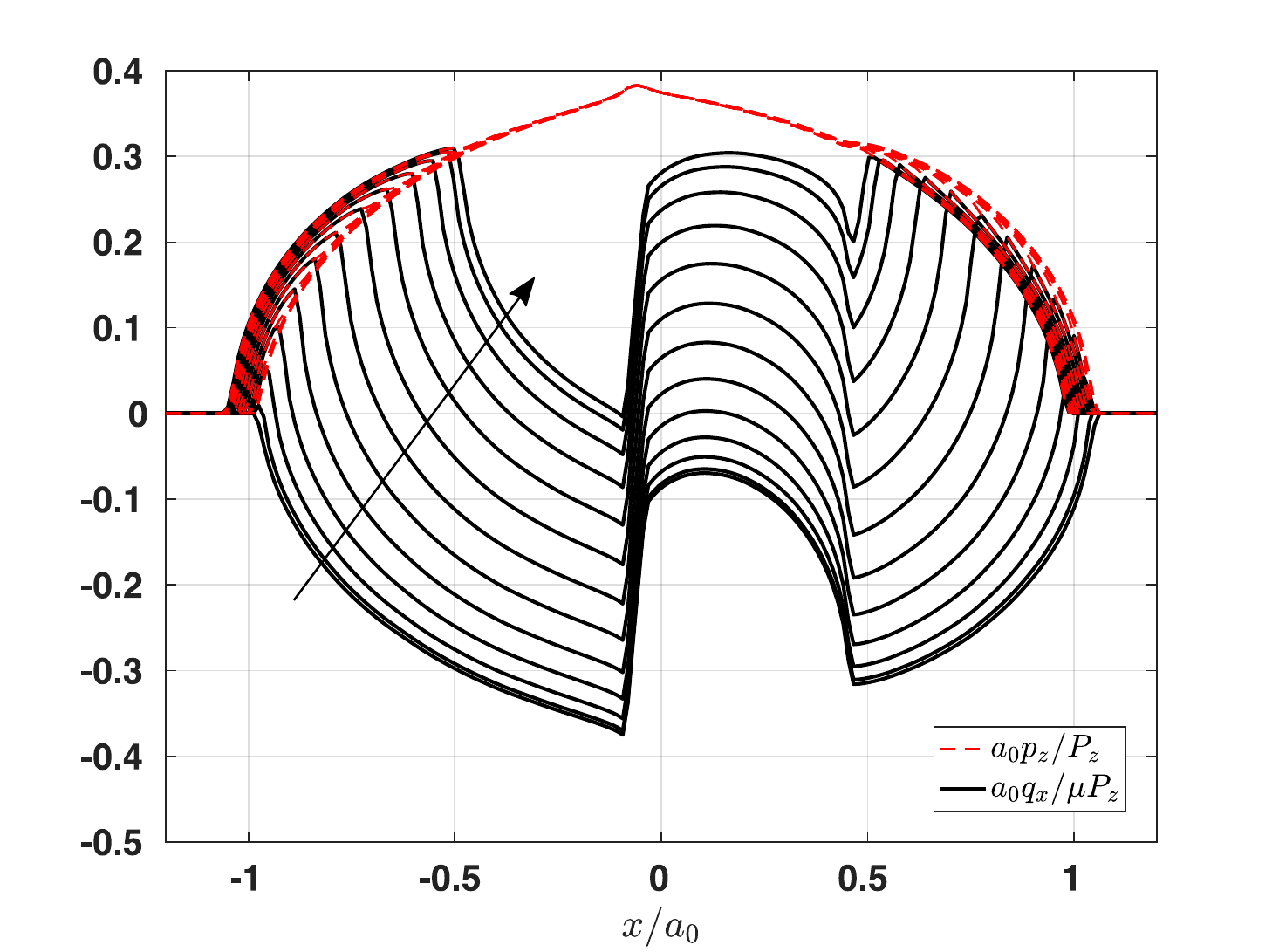}}\\
\caption{tangential cycling load history for low values of coupling, $\beta = 0.29$.}\label{fig:harm2}
\end{figure}

If now the load is reversed, an instantaneous stick zone is created again, which shrinks to a minimum value in correspondence to point $(c)$, Fig.~\ref{fig:harm1c}. The system reaches a stationary state condition after point $(d)$, but with a steady state value of the stick area which is, for a positive load, sensibly different from the one related to the first application of the load, Fig.~\ref{fig:harm1d}. In particular, it retains almost the same extension, but it is shifted towards the center of the contact zone. As a final remark, the system maintains a significant difference between the positive and negative stick areas, even though the steady state is reached after one cycle of loading. It can be noticed that this is directly related to the level of coupling. The current results are obtained with a high level of coupling, corresponding to $\beta=0.50$. The same load history, applied to a system with $\beta = 0.29$ shows greater similarity between the positive and negative steady stick area, Fig.~\ref{fig:harm2}, showing again that the degree of coupling has an important effect on the contact response.
\clearpage
\section{Contact between a rough profile and an elastic layer}\label{sec:5}
\begin{figure}[b!]
\centering
\includegraphics[width=0.5\textwidth]{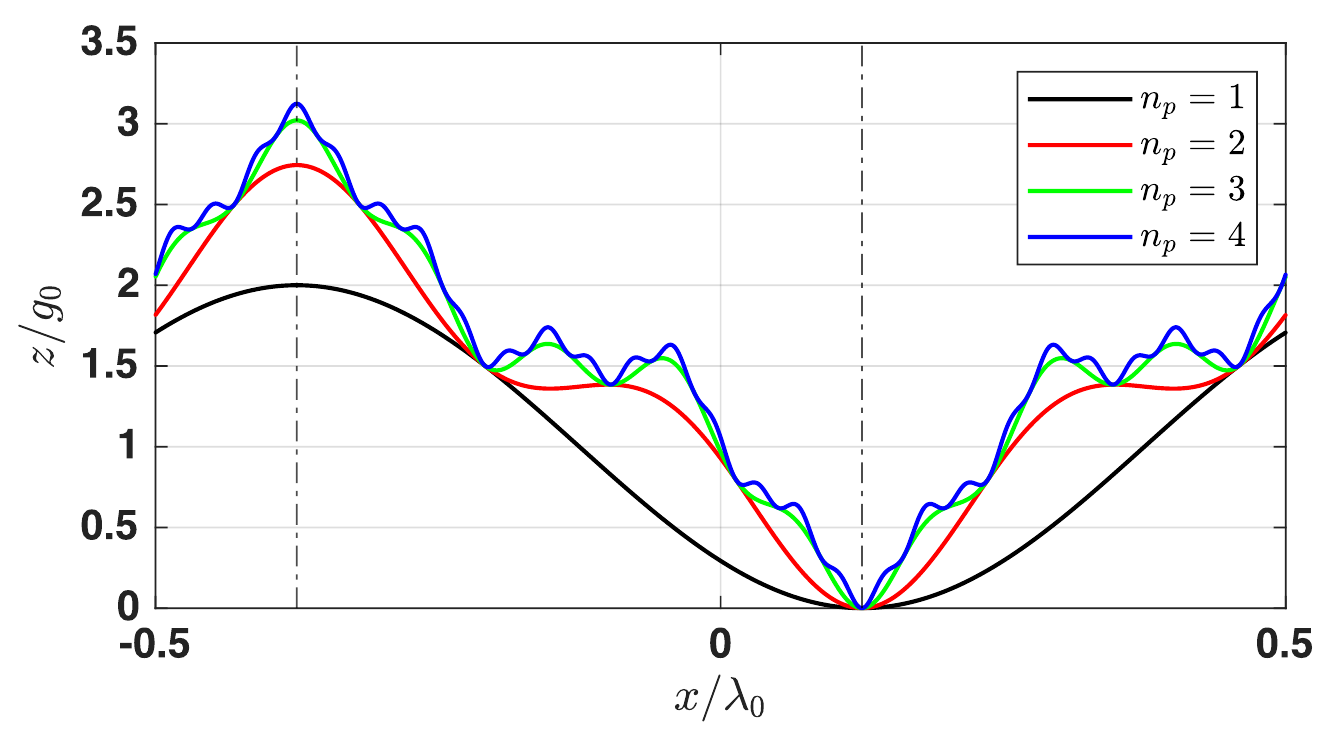}
\caption{Weierstrass multi-scale profile.}
\label{fig:WM_profile}
\end{figure}
As shown in~\cite{MPJR}, the strength of the formulation is particularly evident in the case of contact between a rough profile and an elastic layer. In spite of the fact that any kind of elevation field might be taken into consideration in the computational framework, without any restriction, even numerically generated, a Weierstrass profile is herein used as a possible example. The height field is generated by Eq.~\eqref{eq:WM}:
\begin{equation}\label{eq:WM}
z(x)=g_0\sum_{n=0}^{\infty} \gamma^{(D-2)n}\cos\left(2\pi\dfrac{\gamma^n x}{\lambda_0}\right),
\end{equation}
where in practical applications the summation is carried on up to a certain $n_\mathrm{w}$, thus obtaining a \emph{pre-fractal} profile~\cite{ciavarella,CIAVARELLA20041247}
 which consists of the superposition of $n_\mathrm{w}$ sinusoidal functions, each of them presenting a decreasing wavelength $\lambda_n = \lambda_0/\gamma^n$ and amplitude $g_n = g_0\gamma^{(D-2)n}$, where $\gamma$ and $D$ are parameters chosen such that $\gamma > 1$ and $1\leq D \leq 2$.

In this section, three different indentation problems are solved, in which the contacting profiles exhibit such heights distribution. Each of them is tested against a rectangular elastic block with a height-to-width ratio $t/\lambda_0=0.5$. Such block presents the same elastic parameters employed for the model validation of Sec.~\ref{sec:validation}, i.e. $E_1 = 100$ and $\nu_1 = 0.3$. Each indenter profile can be considered rigid, and is made of the superposition of one, two, or three terms respectively, according to Eq.~\eqref{eq:WM}, shown in Fig.~\ref{fig:WM_profile}. As in the previous section, the lower boundary, i.e. $\partial\mathcal{B}_{\mathrm{C},1}$, is flat, and the elevation field reduces to the one of $\partial\mathcal{B}_{\mathrm{C},2}$: $e_2(\xi) = \bar{h}_2(\xi)-z(\xi)$, with an initial normal gap in the undeformed condition given by $\gnn = \Delta u_\mathrm{n} + z(\xi)$. To obtain an adequate sample of the profile heights field, decreasing characteristics interface mesh sizes have been employed, in accordance with the profiles' shortest wavelength employed from time to time, given as $\lambda_0/\gamma^n$. Following the guidelines of Sec.~\ref{sec:3} results in a number of interface finite elements to be employed, depending on the number of terms of the series of $z(x)$, of $n_\Gamma = \ceil{20\gamma^{n_\mathrm{w}}}$.

Finally, in order to simulate a contact problem which is indefinite in the $x-$direction, periodic boundary conditions have been applied to both the vertical edges of the mesh, at a distance of $\lambda_0$. A classical ironing-type load history is applied for solving the contact problem. First, a purely normal far-field displacement $\Delta_z$ is imposed, starting from zero up to $5g_0$. Then, a horizontal tangential displacement $\Delta_x$ is applied, linearly varying from zero to $3$ times the maximum value of $\Delta_x$, which is the value that guarantees an incipient gross slip for the single harmonics profile. A full parametric study of the problem should involve a thorough evaluation of the sensitivity of the system with respect to the main governing physical parameters, such as $\beta$, $\mu$, $\lambda_0/t$, $g_0$, $\gamma$, $\Delta_x$ and $\Delta_z$, but this is left for further investigation, since the main purpose is to show the feasibility of treating complex interface problems within the present finite element framework.

\subsection{Single harmonics profile in full contact}
\begin{figure}[b!]
\centering
\subfloat[][Purely normal load stage.\label{fig:WM0a}]
{\includegraphics[width=.5\textwidth]{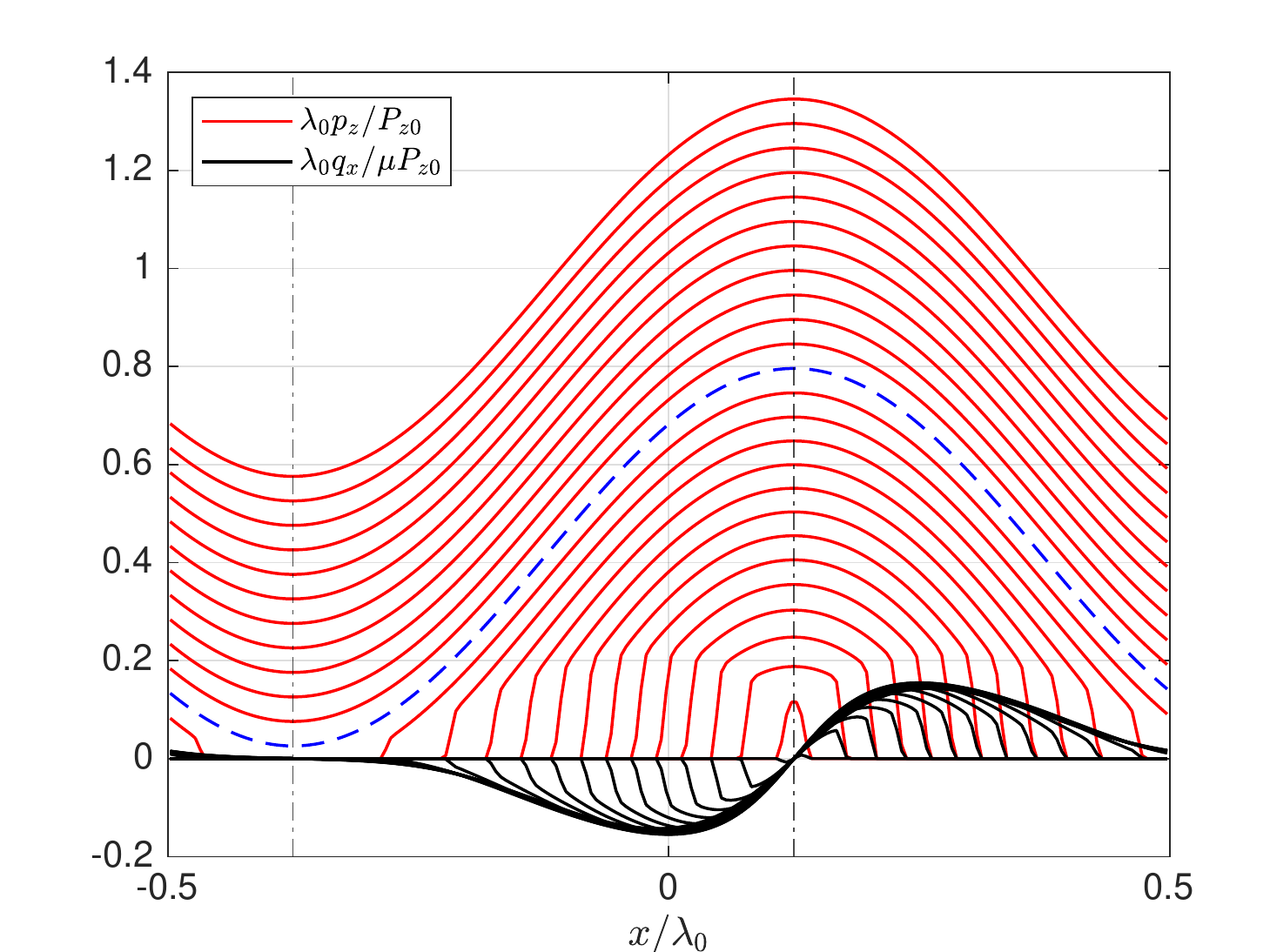}}
\subfloat[][Constant normal load, increasing tangential load stage.\label{fig:WM0b}]
{\includegraphics[width=.5\textwidth]{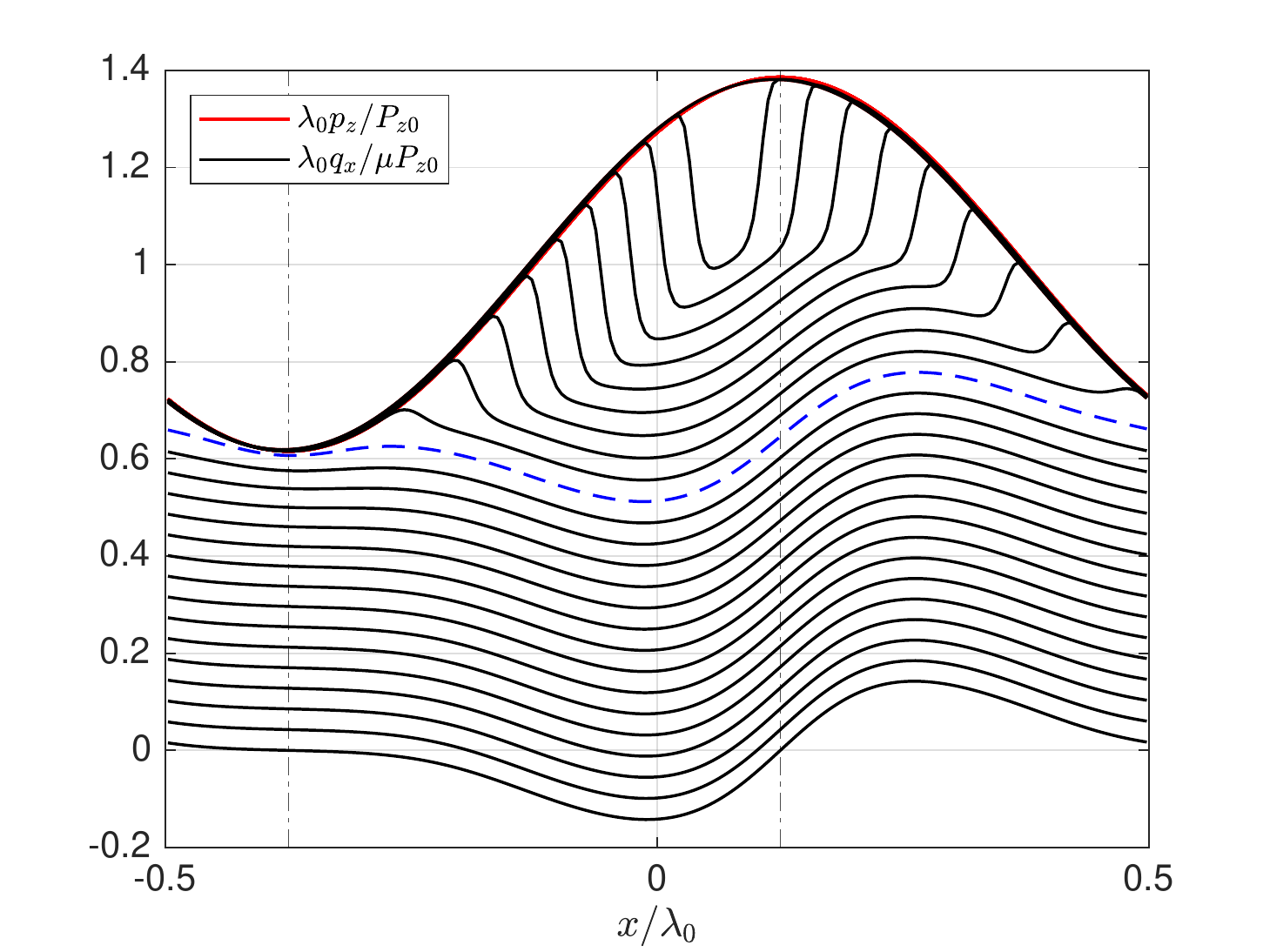}}
\caption{ironing test, single harmonic profile.}\label{fig:WM0}
\end{figure}
The case of $n_\mathrm{w}=0$ is shown in Fig.~\ref{fig:WM0}. The main difference with the results obtained for the previous test problems is that now we are dealing with an infinitely long profile which makes contact at an infinite set of spots. Under purely normal loading, the vertical tractions $p_z(x)$ present two axes of symmetry, highlighted by dash-dotted lines in the figure, which corresponds to axes of anti-symmetry for the shearing tractions $q_x(x)$. Since in this condition shearing tractions must be strictly null in each of these points, they result in having a lower intensity, and the absence of the characteristic backward and forward slip zones which are typical of the Hertzian problem. 

Tangential tractions grow in intensity and extension until the full contact condition is reached, corresponding to a value of $p_z(x)$ which is highlighted by the blue dashed curve in Fig.~\ref{fig:WM0a}. 
After that point, since a full stick condition holds, $q_x(x)$ remains constant until the maximum value of the vertical far-field displacement is reached. After that point, the horizontal far field displacement is applied, and the horizontal tractions grow until a condition of partial slip is reached, again blue dashed line in Fig.~\ref{fig:WM0b}. As expected, the last point of the interface coming into contact is also the first one which undergoes partial slip. After this point, the state of the system is such that there is an alternation of shrinking stick islands bordered by increasing zones of full slip. When the transient regime ends, a perfect overlapping between $\mu p_z(x)$ and $q_x(x)$ is observed.

\subsection{Multiple harmonics profile contact}
\begin{figure}[b!]
\centering
\subfloat[][Purely normal load stage.\label{fig:WM1a}]
{\includegraphics[width=.5\textwidth]{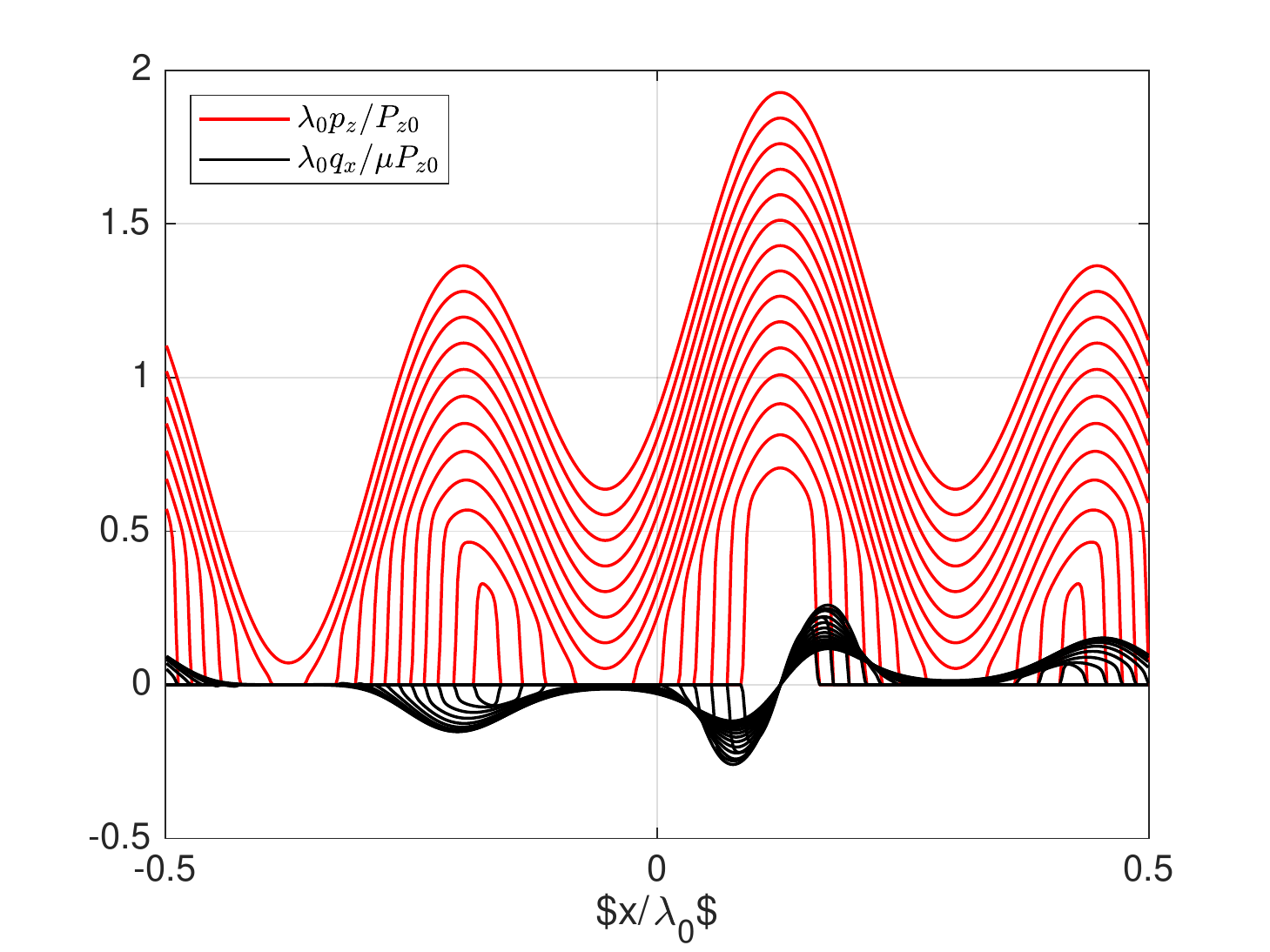}}
\subfloat[][Constant normal load, increasing tangential load stage.\label{fig:WM1b}]
{\includegraphics[width=.5\textwidth]{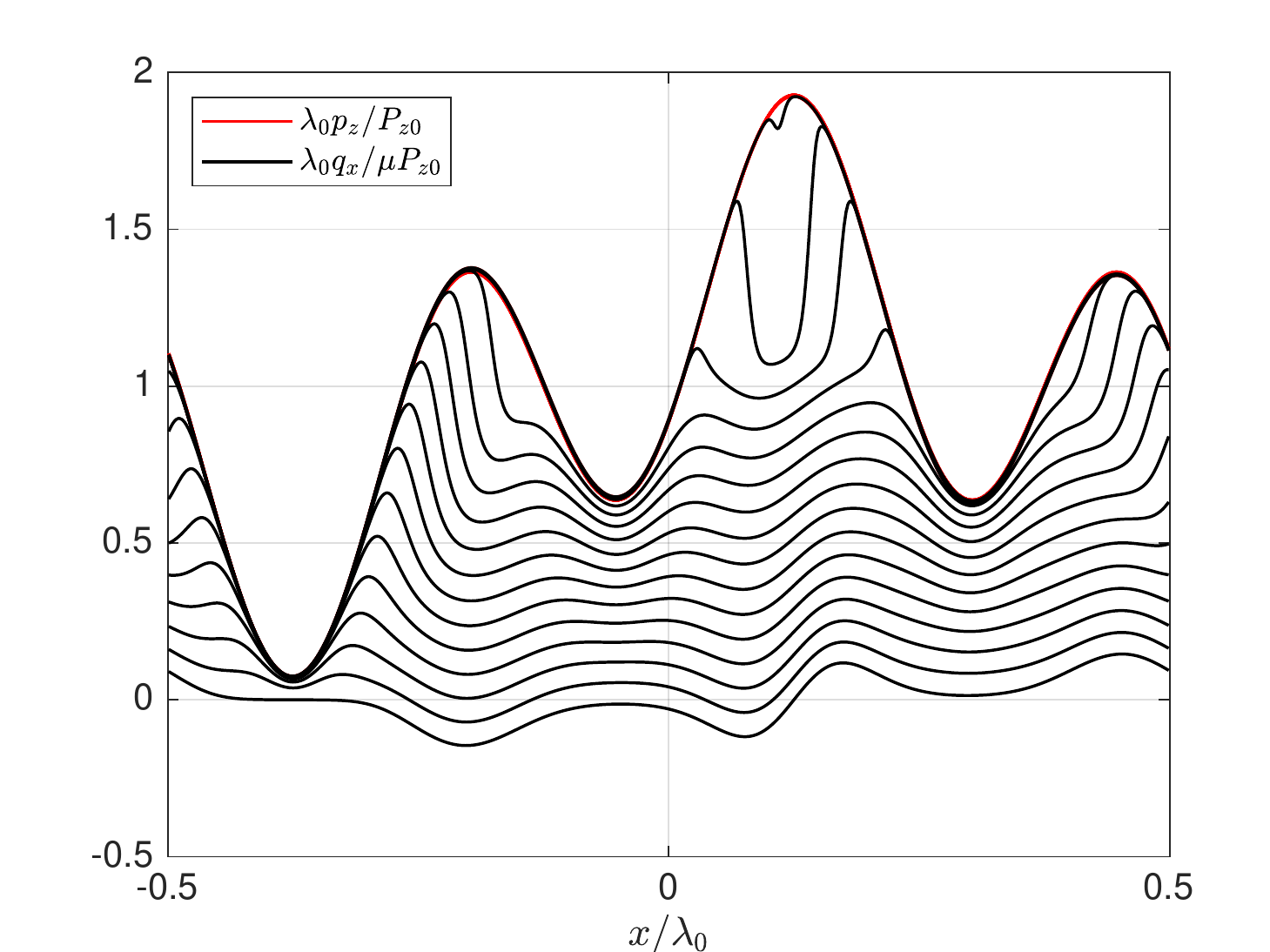}}
\caption{ironing test for double harmonics profile.}\label{fig:WM1}
\end{figure}
The addition of a length-scale in the Weierstrass function has the immediate effect of increasing the peak values of the normal tractions, which are localised in correspondence of the local maxima of the profile, and of reducing the contact area for a given level of the external load: the full contact condition is still achieved, but after a higher number of time steps, see Fig.~\ref{fig:WM1a}. The same considerations on the distribution of the tangential tractions can be made also in this case, with the difference that now the contact domain is no longer compact.

In this case, an approximated study of $q_x(x)$ and of the extension of the stick and slip areas can still be possible, exploiting, e.g the Ciavarella-J{\"a}ger theorem, but under the limiting assumptions of uncoupling and equality of the Green functions in $x$ and $z$ directions. An interesting feature that could be appreciated thanks to having taken into account coupling effects is that, still under purely normal loading, if two asperities, each of them generating a separate contact island, are characterised by a severe gradient in terms of vertical tractions, the less pressed one might experience a horizontal traction distribution which is very far from the one typical of an isolated asperity, i.e. anti-symmetric. If the vertical tractions in the leading asperity are high enough, the horizontal displacements generated by them might be so high as compared to the ones generated by the secondary asperity that the latter are negligible, thus resulting in horizontal tractions which are all negative or positive valued from the beginning. With suitable boundary conditions or values of $\mu$, there might also be a condition of gross slip from the beginning of the contact process.
\begin{figure}[h!]
\centering
\subfloat[][\label{fig:close1}]
{\includegraphics[width=.5\textwidth]{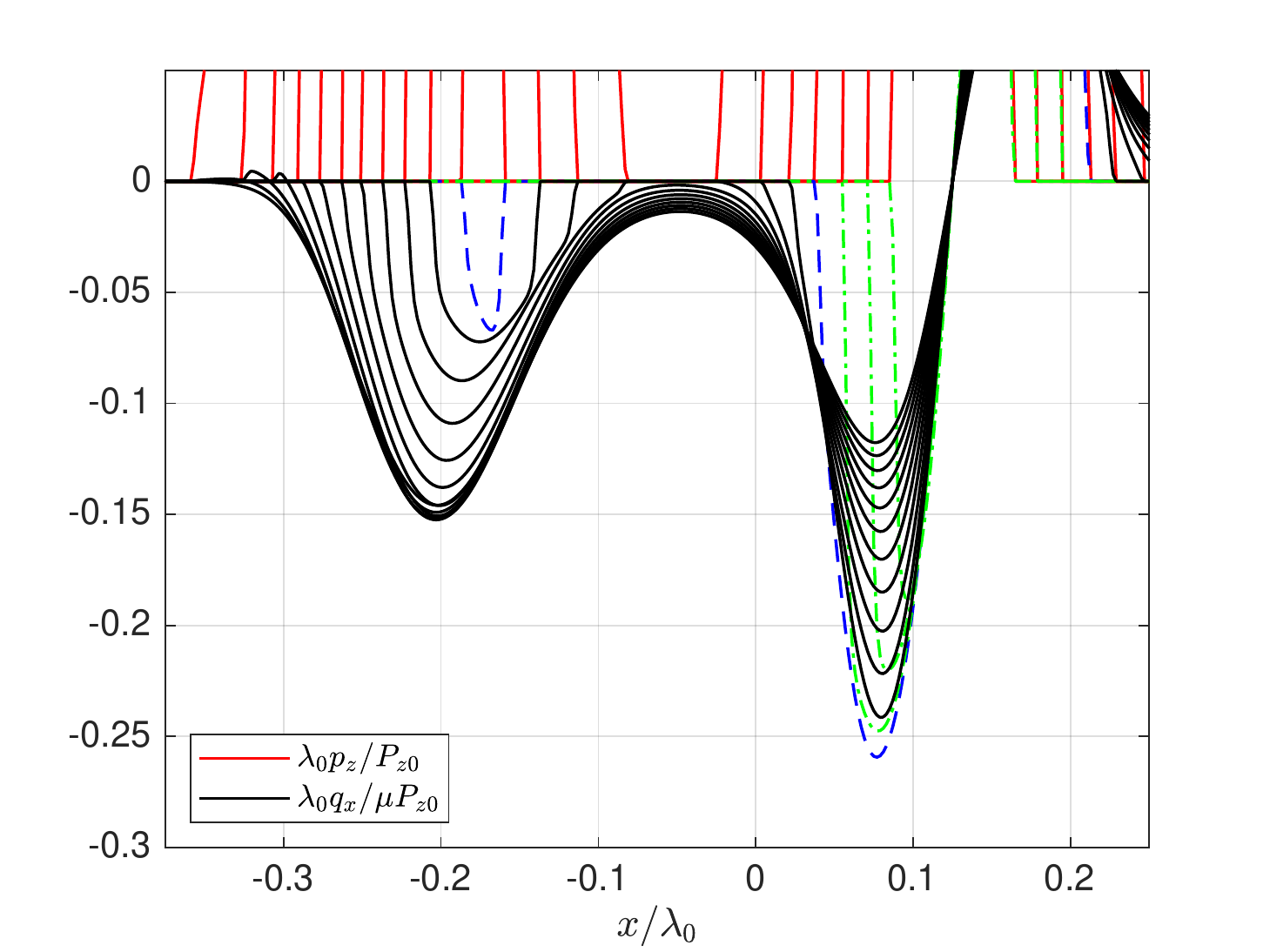}}
\subfloat[][\label{fig:close2}]
{\includegraphics[width=.5\textwidth]{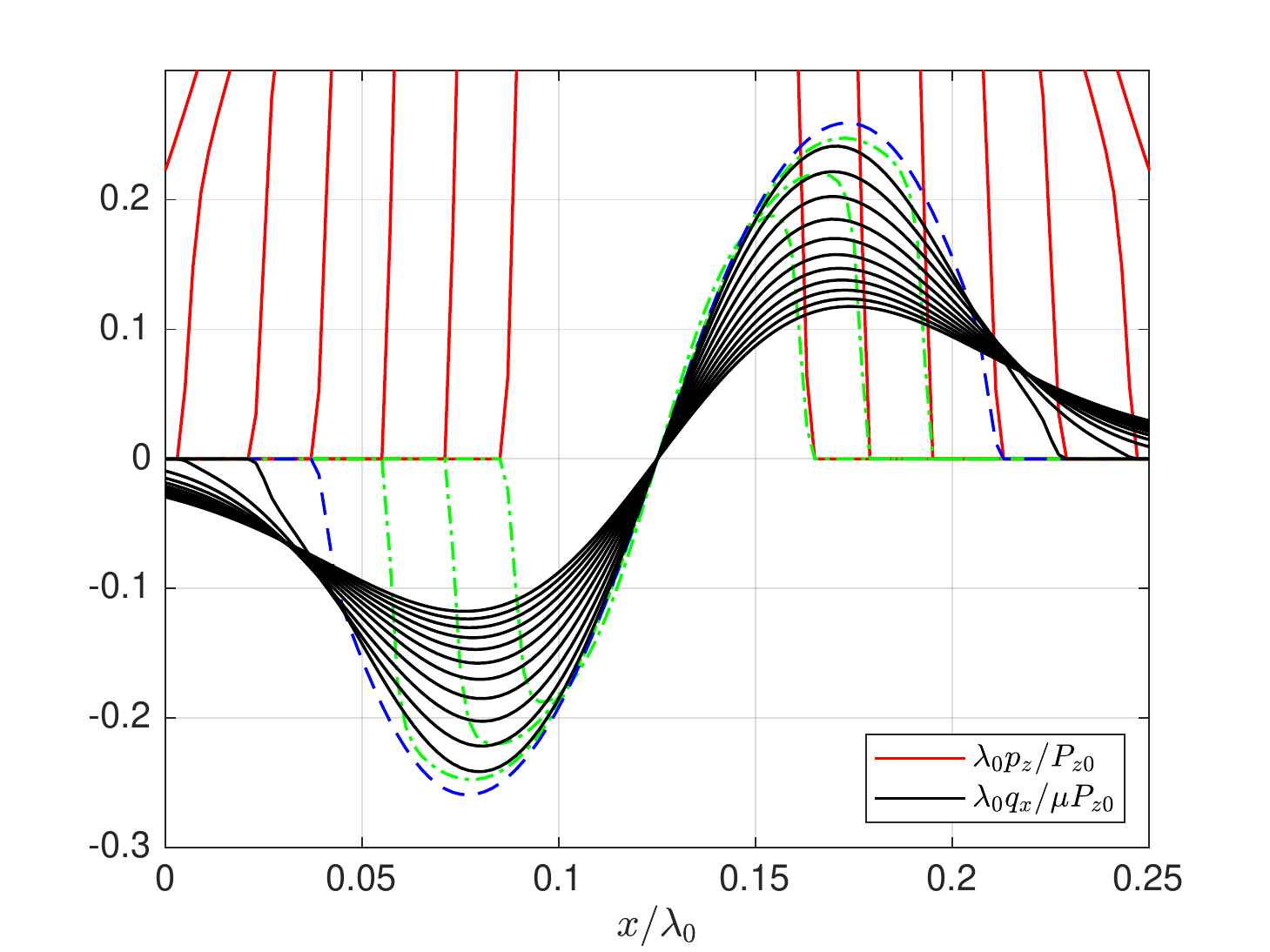}}
\caption{Evolution of contact tractions.}\label{fig:close}
\end{figure}
At the same time, when the second asperity comes into contact, it exerts a stiffening effect on the bulk, which reflects in a decrease of the magnitude of the increment rate of the horizontal displacements towards the high asperity, which in the final place determines a relaxing of the tangential tractions at the level of the leading asperity. This characteristics is depicted in Fig.~\ref{fig:close1} and~\ref{fig:close2}. The tangential tractions over the leading asperity, green dash-dotted curves, increase in extension and magnitude as long as the second asperity comes into contact, where they reach their maximum value, blue dashed line. After that moment, they continue growing in extension, since the contact area is still increasing, but they decrease in magnitude, due to the interaction between the different contact islands. Finally, when also the tangential far field displacement is applied, they start growing in magnitude again, gross slip starts, and the transition between full stick and full slip takes place, see Fig.~\ref{fig:WM1b}.
\begin{figure}[h!]
\centering
\subfloat[][Purely normal load stage.\label{fig:WM2a}]
{\includegraphics[width=.5\textwidth]{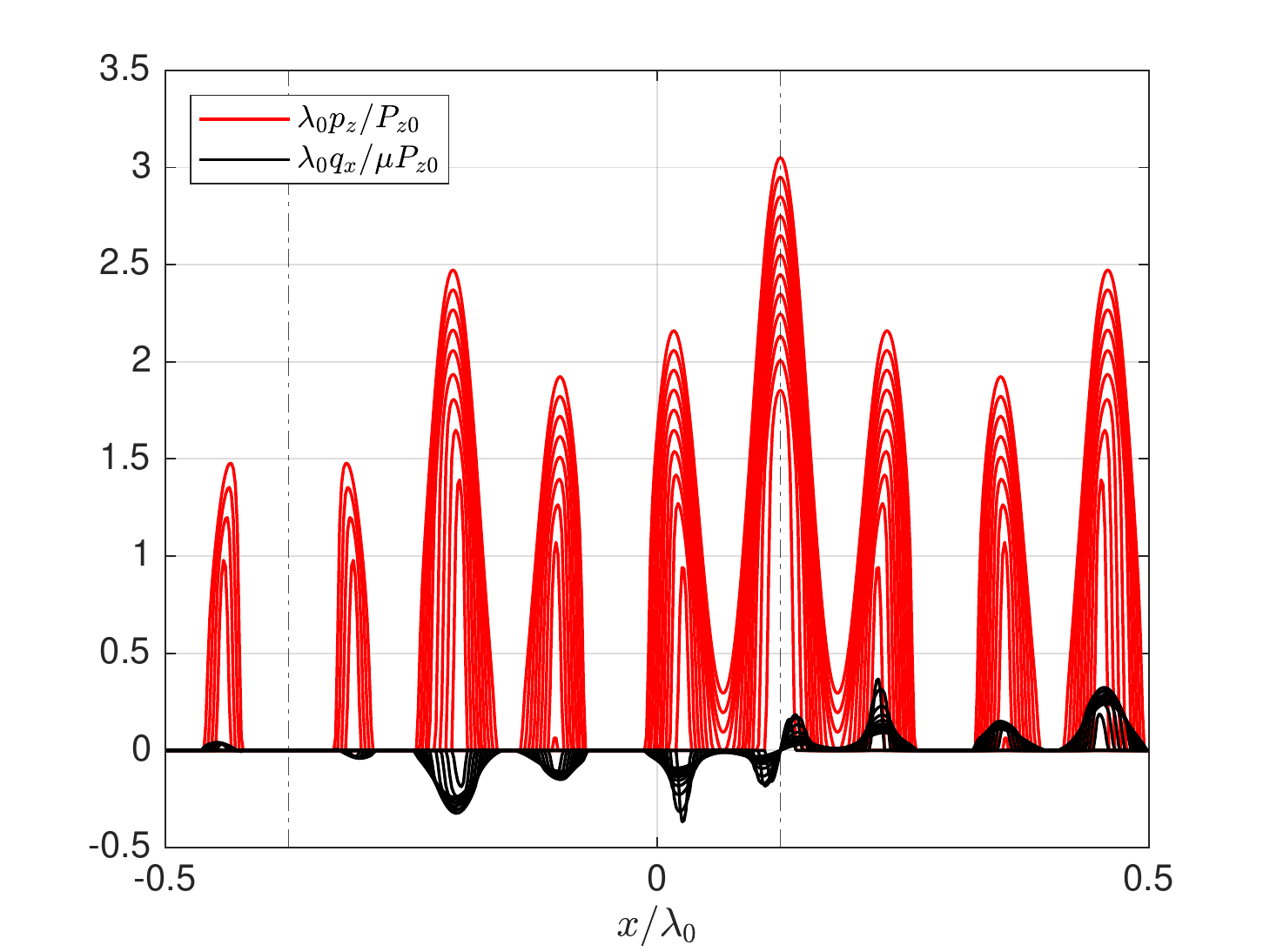}}
\subfloat[][Constant normal load, increasing tangential load stage.\label{fig:WM2b}]
{\includegraphics[width=.5\textwidth]{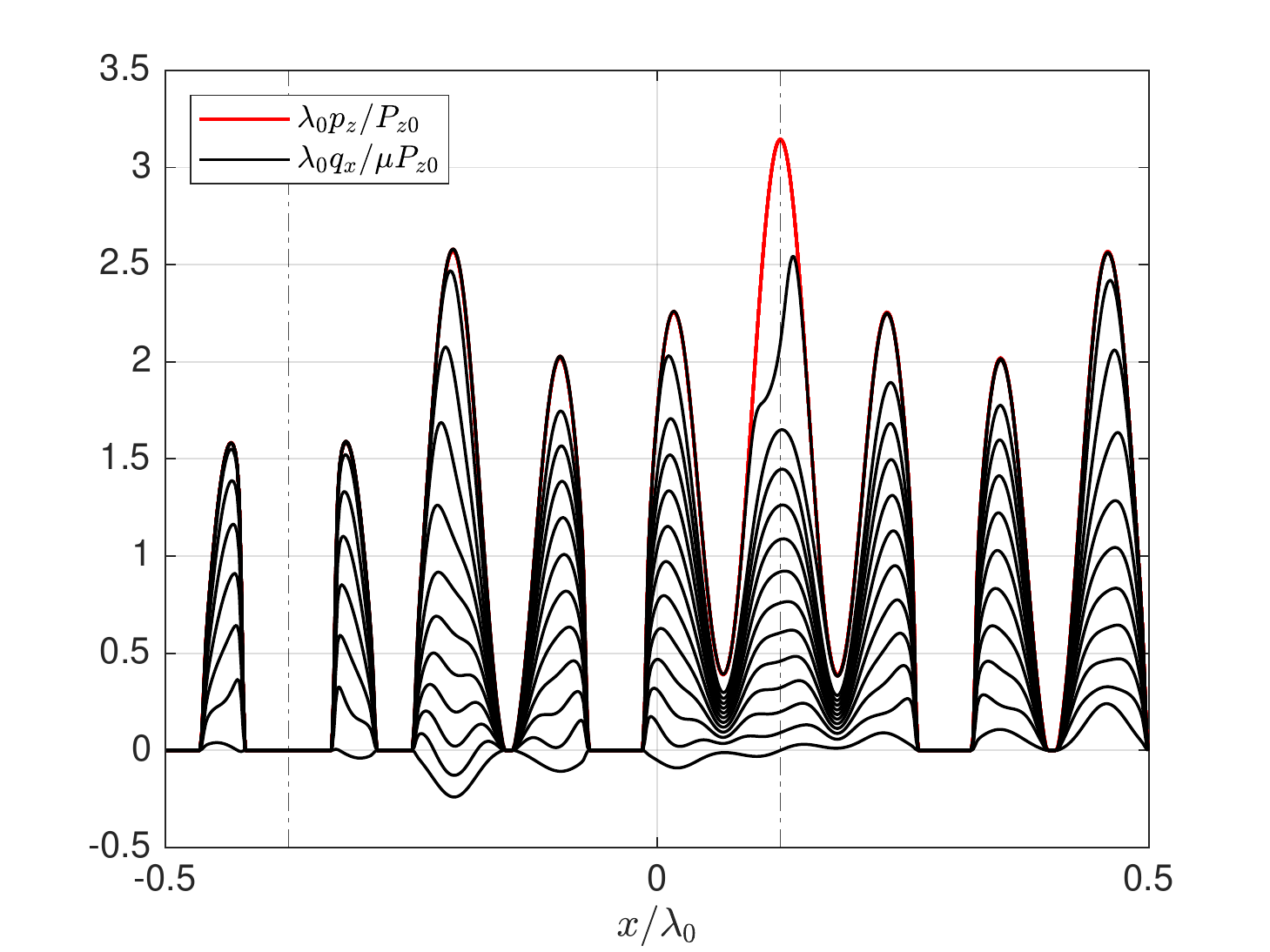}}
\caption{ironing test for a triple harmonics profile.}\label{fig:WM2}
\end{figure}
The same trend can be observed also for the profile characterised by $n_\mathrm{w}=2$, Fig.~\ref{fig:WM2}, where the same comments apply as well for the increase in vertical tractions, the reduction of the contact area and the evolution of the stick and slip zones.
\section{Discussion and conclusion}
The proposed formulation provides a way to overcome the shortcomings and the difficulties which are encountered during the solution of contact problem between rough or generically complex profiles, when friction is considered. A standard approach of explicitly modeling the geometry of the interface using higher order interpolation schemes such as Bezier curves, adaptive mesh refinement, or NURBS, is more suitable when the interface consists of regular and smooth profiles, while it can be difficult to exploit or it could be very expensive from a computational point of view when rough profiles have to be analysed.

In this work, the comprehensive and challenging  extension to frictional contact problem of the framework which has been set up in~\cite{MPJR} (denominated as  eMbedded Profile for Joint Roughness (MPJR interface finite element)) has been proposed. The fundamental idea was to re-cast the original geometry of the contacting profiles, obtaining a macroscopically smooth interface which allows for a straightforward meshing with linear finite elements, while the actual geometry is stored in terms of its analytical expression and passed to the system as a correction to the initial gap function, thanks to the assumption of having a rigid indenting profile. The major advantages obtained by exploiting this approach are the use of a low order finite element interpolation scheme, with a significant reduction of nodal degrees of freedom and save of computational time.

The classical benchmark tests which are usually employed for testing the capability of higher order interpolation schemes have been used for validating the model, with excellent results obtained even for relatively coarse interface discretisations. Finally, the method has been successfully tested in relation to more complex scenarios of contact problems involving a Weierstrass profile with multiple harmonics, resulting in a useful tool for the investigation of the behavior of idealised 2D fractal rough surfaces under the non trivial assumption of full coupling between normal and tangential traction fields.

The natural development of the presented interface element, which by itself stems from~\cite{MPJR}, includes its extension to three dimensions and possibly the inclusion of other interface phenomena which could be much more difficult to be analyzed using standard FEM or BEM approaches, such as the interplay of friction and adhesion, or friction and plasticity. The presented implementation  also allows for the possibility of more complex friction laws to be used, as, for example, the ones employed in~\cite{REZAKHANI2020103967}.

Results have highlighted the important role of coupling between normal and tangential contact problems, with a special focus on rough surfaces, which is an open research topic also for precision engineering applications.

Finally, forthcoming studies would encompass the corresponding  formulation to geometrically nonlinear effects and prospective coupling of contact-induced fracture events. Such developments are beyond the scope of the present investigation, deserving a  careful attention.

\section*{Acknowledgements}
Authors would like to acknowledge funding from the MIUR-DAAD Joint Mobility Program 2017 to the project "Multi-scale modeling of friction for large scale engineering problems". The project has been granted by the Italian Ministry of Education, University and Research (MIUR) and by the Deutscher Akademischer Austausch Dienst (DAAD). The corresponding author would also like to acknowledge the contribution of the Erasmus+ European project that supported his mobility to Bundeswehr Univerit{\"a}t M{\"u}nchen (Germany), where he focused on the study of frictional contact problems. 

MP would like to acknowledge the support from the Italian Ministry of Education, University and Research (MIUR) to the Research Project of Relevant National Interest (PRIN 2017) XFAST-SIMS: Extra-fast and accurate simulation of complex structural systems.

JR   is thankful  to  the Consejer\'ia de Econom\'ia y Conocimiento of the Junta de Andaluc\'ia (Spain)  contract US-1265577-Programa Operativo FEDER Andaluc\'ia 2014-2020 and the Ministerio de Ciencia, Innovación y Universidades  (Spain) contract  PID2019-109723GB-I00.

\section*{Declaration of interests}
The authors declare that they have no known competing financial interests or personal relationships that could have appeared to influence the work reported in this paper.


\bibliography{ref}

\end{document}